\documentclass[12pt]{article}
\usepackage{cite}
\usepackage{graphicx}
\usepackage{subfigure}
\usepackage{amssymb,amsmath}
\usepackage{psfrag}

\begin{document}

\title{\Large Phenomenological Constraints on Extended Quark Sectors}
\author{Tomer Yanir\footnote{e-mail: tomer.yanir@weizmann.ac.il}\\
  \small {Department of Particle Physics, Weizmann Institute of
    Science, Rehovot 76100, Israel}}
\date{}
\maketitle

\begin{abstract}
  We study the flavor physics in two extensions of the quark sector of the
  Standard Model (SM): a four generation model and a model with a
  single vector--like down--type quark (VDQ).
  In our analysis we take into account the experimental constraints from
  tree--level charged current processes, rare Kaon decay processes,
  rare $B$ decay processes, the $Z\to b \bar{b}$ decay,
  $K$, $B$ and $D$ mass differences, and the CP violating
  parameters $\frac{\varepsilon^\prime}{\varepsilon}$, $\varepsilon_K$
  and $a_{\psi K}$. All the constraints are taken at two sigma.
  We find bounds on parameters which can be used to represent the
  New Physics contributions in these models ($\lambda_{t^ \prime}^{bd}$,
  $\lambda_{t^ \prime}^{bs}$ and $\lambda_{t^ \prime}^{sd}$ in the
  four--generation model, and $U_{bd}$, $U_{bs}$ and $U_{sd}$ in the
  VDQ model) due to all the above constraints.
  In both models the predicted ranges for $a_{SL}$
  (the CP asymmetry in semi-leptonic decays), $\Delta M_D$, $B(K^+\to\pi^+ \nu \bar{\nu})$,
  $B(K_L\to\pi^0 \nu \bar{\nu})$ and $B(K_L\to \mu \bar{\mu})_{SD}$ can be
  significantly higher than the predictions of the SM, while
  the allowed ranges for $a_{\psi K}$ and for
  $\Delta m_{B_S}$ are consistent with the SM prediction.
\end{abstract}


\section{Introduction}
\label{sec:introduction}

Extensions of the quark sector modify many features of the
Standard Model flavor physics. In particular, CKM unitarity is
violated, and there are new sources of flavor--changing
neutral--currents (FCNC) and of CP violation. We here analyze the
flavor physics of two such extensions: a fourth generation of
fermions and a single vector--like representation of quarks added
to the SM.

These two models share many common aspects. Both models have the
same number of mixing angles and phases, which can be taken as
parameters of a $4\times4$ unitary matrix. Both models violate the
$3\times 3$ CKM unitarity in a similar manner, and introduce new
sources of CP violation. Although originating from different
Feynmann diagrams, many of the expressions for the experimental
observables in the two models are practically the same (except for
numerical differences). The features of CKM non-unitarity and new
CP violating parameters are of particular interest, since they
offer us a behavior which is qualitatively different from that of
the SM. Also, the two models provide new contributions to FCNC
processes. Since FCNC are highly suppressed in the SM, they
provide a useful tool in searching for New Physics and in
constraining it.

In order to constrain the models, we consider the new
contributions to rare $K$ and $B$ decays, neutral meson mixing
mass differences, CP violating parameters in the $K$ and in the
$B$ systems, and the process $Z \to b \bar b$. To obtain the
numerical results, we scan the allowed ranges of the mixing
parameters. For each point in this parameter space, we check that
all the above mentioned constraints are obeyed. This gives us
bounds on the possible values of the mixing parameters and the
correlations between them. We also use the scan to give
predictions for various observables in these models, and compare
them to the predictions in the SM. Finally, we analyze the
predictions of these models to various processes that are not
measured yet, such as $K^+\to\pi^+ \nu \bar{\nu}$, $K_L\to\pi^0
\nu \bar{\nu}$,  $B\to X_s \ell^+ \ell^-$, $\Delta M_D$ and
$\Delta M_{B_S}$.

The paper is organized as follows. In
section~\ref{sec:four_gen_model} we describe the analysis for the four generation
model, and in section \ref{sec:vldt} we describe the
analysis for the VDQ model. Each of these sections includes
some background about the model, descriptions of the various constraints
used in the analysis, and the details of the numerical results.
Finally, in section~\ref{cha:discussion-conclusions} we discuss our
conclusions.


\section{A four generation model}
\label{sec:four_gen_model}


\subsection{Background}
\label{sec:background}

Currently there is no known fundamental principle which fixes the
number of SM generations. Regarding experimental constraints, as
of today there is no conclusive evidence that excludes a fourth
generation. There are, however, two problematic issues. First, the
invisible decay width of the $Z^0$ boson clearly indicates the
existence of exactly three light neutrinos. The existence of a
fourth neutrino with mass $m_{\nu4} \ge 45$ GeV is not excluded by
the data, but it requires some mechanism that will give the new
neutrino a large mass while keeping the masses of the SM neutrinos
small. Second, the four--generation model has some difficulties in
explaining the electroweak precision measurements. Various
analyzes of electroweak precision measurements (e.g.
\cite{PDG_2000} and \cite{Ilyin_2000}) differ in their conclusions
regarding the implications for a fourth generation. We assume that
these measurements are consistent with four generations.
Specifically, our analysis can be viewed as complementary to the
one in ref. \cite{Ilyin_2000}, taking into account also the mixing
in the quark sector.

The constraints we consider are the following: charged--current
tree--level decays, the branching ratios $B(K^+\to\pi^+\nu\bar{\nu})$,
 $B(K_{L}\to\mu\bar{\mu})_{SD}$ and $B(B\to X_s \ell^+ \ell^-)$, the mass differences $\Delta m_K$,
$\Delta m_{B_d}$, $\Delta m_{B_s}$ and $\Delta m_{D}$, the CP violating parameters
$\varepsilon_K$ and $a_{\psi K}$, and the partial decay width of $Z\to b \bar{b}$.

An extensive work on this subject can be found in \cite{Hattori_1999}.
We add to it the $B\to\psi K_{S}$ constraint, which was not available
at the time, and the $Z\to b \bar{b}$ constraint. Also, we treat
$B(B\to X_s \ell^+ \ell^-)$ in a more careful manner.
In addition, we update the experimental bounds used in
\cite{Hattori_1999} and we consider the entire range of possible mixing
angles and phases.
Another work which considered the tree--level
decays, a rough analysis of the $Z\to b \bar{b}$ process and the $B$ meson system
constraints, can be found in \cite{Eyal_1999}.

\subsection{The model}
\label{sec:four_gen_model_specifics}

We consider a model with an extra generation of chiral fermions added to
the SM in the simplest possible way. The representations of the fourth
generation fermions are identical to the representations of the usual three SM generations.
In such a model, there are nine additional parameters compared to the SM:
four new masses, and five new mixing angles and phases. The origin of
the new mixing angles and phases
comes from the fact that the CKM matrix is now a $4\times4$ matrix.
This means that it consists of nine physical parameters:
six mixing angles (compared to three in the SM) and three complex
phases (compared to one in the SM). We use a specific parametrization
of the $4\times4$ CKM matrix~\cite{Botella_1986}, in order to
incorporate all the correlations in the analysis which we perform.
The mixing angles will be referred to as $\theta_{12}, \theta_{13},
\theta_{23}, \theta_{14}, \theta_{24}$ and $\theta_{34}$,
and the phases as $\phi_{13}, \phi_{14}$ and $\phi_{24}$.
In the limit of vanishing new mixing angles, one gets the
usual mixing angles $\theta_{12}, \theta_{13}, \theta_{23}$ and the CP
violating phase $\phi_{13}$ of the SM.

We deal only with the quark sector of the model, which includes the SM
quarks as well as the new quarks (to be referred to as
($t^\prime\,\, b^\prime$)). The new leptons do not play any role in
our analysis, and we do not discuss them. In particular, we do
not restrict ourselves to any specific
mechanism that gives rise to the high mass required for the fourth neutrino.

The effects of the fourth generation enter the processes we consider through loop
diagrams: the new quarks $t^\prime$ and $b^\prime$ can now appear in
the loops. In most of the constraints, the fourth generation
contribution enters almost exclusively through the parameters $\lambda_{t^\prime}^{bd}$,
$\lambda_{t^\prime}^{bs}$ and $\lambda_{t^\prime}^{sd}$,
where
\begin{equation}
  \lambda_k^{lm}\equiv V_{kl}^*V_{km}.
  \label{eq:lambda_defined}
\end{equation}
These parameters can be used to evaluate the effects of New Physics contributions.


\subsection{The Constraints}
\label{sec:the-constraints}

The analyzes done in \cite{PDG_2000,Ilyin_2000} concerning
electroweak precision measurements with four generations indicate
that only a restricted range of $m_{t^\prime}-m_{b^\prime}$ is
allowed:
\begin{equation}
  |m_{t^\prime}-m_{b^\prime}| \leq 85 \ \rm{GeV\ \ \ \ (at\ 95\%
   \ CL)}.
  \label{eq:delta_mbtprime_constraint}
\end{equation}
Another constraint on the masses comes from the combined results
of direct measurements at CDF and
D0 collaborations~\cite{CDF_bprime__2000}, which rule out
$m_{b^\prime} \leq 175$ GeV, assuming that the FCNC decay mode is
dominant. We assume hereafter $m_{t^\prime} \geq
m_{b^\prime} \gtrsim 175$ GeV. Also, since a perturbative approach
is assumed to be valid, one must require $m_{t^\prime}\,
,\,m_{b^\prime} \lesssim 500$ GeV.
In view of the above constraints, we consider in our analysis two extreme cases of the
new fermion masses. The first uses the highest possible mass
values (in view of the above constraints): $m_{t^\prime}=500$ GeV,
$m_{b^\prime}=470$ GeV. The second uses $m_{t^\prime}=200$ GeV,
$m_{b^\prime}=170$ GeV.
For each of the two cases ($m_{t^\prime}=500$ GeV or $m_{t^\prime}=200$
GeV), any alternative choice for $m_{b^\prime}$ (which obeys the
limitation given in eq. \eqref{eq:delta_mbtprime_constraint}) can
cause only minor changes to the results we present in this work.
The actual calculations for both cases ($m_{t^\prime}=500$
GeV and $m_{t^\prime}=200$ GeV) are basically the same, with only
numerical differences.
In order not to repeat the presentation
twice, we give a detailed analysis only
for the case $m_{t^\prime}=500$ GeV. The results
for the second case (with $m_{t^\prime}=200$ GeV) are briefly
summarized in section \ref{sec:mtp_200}.

Measured SM tree--level processes are not
affected by the existence of a fourth generation, and can
therefore be used as in the SM to constrain elements of the CKM
matrix. This enables us to bound the magnitude of the matrix
elements $V_{ud}, V_{us}, V_{ub}, V_{cd}, V_{cs}$ and $V_{cb}$.
\begin{table}[htb]
\begin{center}
\footnotesize{
\begin{tabular}{|c|c|c|c|}
\hline
Parameter & Mean value & Sigma & Ref. \\
\hline
$\left | V_{ud} \right |$  & $0.9735$    & $ 0.0008 $ & \cite{PDG_2000}\\
$\left | V_{us} \right |$  & $0.2205$    & $ 0.0018 $ & \cite{Buras_2001}\\
$\left | V_{ub} \right |$  & $0.00349$   & $ 0.00076 $ & \cite{Buras_2001}\\
$\left | V_{cd} \right |$  & $0.224$     & $ 0.016 $ & \cite{PDG_2000}\\
$\left | V_{cs} \right |$  & $0.996$     & $ 0.024 $ & \cite{CKM_Bargiotti_2001}\\
$\left | V_{cb} \right |$  & $0.041$     & $ 0.002 $ &
\cite{Buras_2001}\\
$R_b$                      & $0.21653$   & $ 0.00069 $ & \cite{LEP_2001}\\
$Br(K_L\to\mu\bar\mu)_{SD}$& $\leq 3.75\times 10^{-9} $  & 95$\%$ CL & \cite{AMBROSIO_1998,E871_2000}\\
$Br(K^+\to\pi^+\nu\bar\nu)$ & $\leq 5.07\times 10^{-10} $ & 95$\%$ CL & \cite{E787_2001}\\
$Br(K_L\to\pi^0\nu\bar\nu)$ & $\leq 5.9\times 10^{-7} $  & 95$\%$ CL & \cite{Buras_2001}\\
$B(B\to X_s e^+ e^-)$      & $\leq 1\times 10^{-5} $  & 90$\%$ CL & \cite{ABE_2001}\\
$\varepsilon_K$            & $2.28\times 10^{-3}$       & $0.013\times 10^{-3}$ & \cite{Buras_2001}\\
${\varepsilon^\prime}\over\varepsilon$ & $17.2\times 10^{-4}$ &
$1.8\times 10^{-4}$ & \cite{NA48_2001}\\
$\Delta m_K$               & $3.489\times 10^{-15}$ GeV & $0.008\times
10^{-15}$ GeV& \cite{Buras_2001}\\
$\Delta m_{B_d}$           & $3.2\times 10^{-13}$ GeV   & $0.092\times 10^{-13}$ GeV& \cite{Buras_2001}\\
$\Delta m_{B_s}$           & $> 9.87\times 10^{-12}$ GeV   & 95$\%$ CL & \cite{Buras_2001}\\
$a_{\psi K}$             & 0.8                      & 0.1 & \cite{BABAR_2001_apsi,BELLE_2000,CDF_1999}\\
$|M_{12}^D|$               & $\leq 6.3\times 10^{-14}$ GeV & 95$\%$ CL & \cite{RAZ_2001}\cr
\hline
\end{tabular} }
\caption{ \small { Values of experimental-related input parameters used in the analysis. }}
\label{tab:experimental_data}
\end{center}
\end{table}
The parametrization of $V_{CKM}$ means that we can translate these
constraints into bounds on some of the mixing angles. Taking the
constraints as described in Table~\ref{tab:experimental_data} at
two sigma, and using unitarity of the $4\times4 \, V_{CKM}$
matrix, we deduce the following ranges:
\begin{eqnarray}
  \label{eq:mixing_bounds}
  &\sin\theta_{12} = 0.22 \ , \nonumber\\
  &\sin\theta_{23} = 0.041\pm 0.004 \ , \nonumber\\
  &\sin\theta_{13} = 0.00349 \pm 0.00152 \ ,\\
  &0 \leq \sin\theta_{14} \lesssim 0.085 \ ,  \nonumber\\
  &0 \leq \sin\theta_{24} \lesssim 0.25 \ . \nonumber
\end{eqnarray}
These bounds are independent of the new quark masses. The
parameter $\sin\theta_{34}$ and the CP violating phases $\phi_{13},
\phi_{14}$ and $\phi_{24}$ remain unconstrained at this stage.

We now consider loop processes. These depend on the Inami-Lim
functions \cite{InamiLim_1981} $X_0(x)$, $Y_0(x)$ and $S_0(x)$ defined
in \cite{Buras_1996}. Experimental inputs are given in table
\ref{tab:experimental_data} and theoretical ones in table \ref{tab:theoretical_data}.
Most of the expressions for the various constraints in
the four generation model are obtained in a simple manner from the
relevant expressions for the SM, by considering
additional diagrams with $t^\prime$ quarks in the loop.
\begin{table}[htb]
\begin{center}
\footnotesize{
\begin{tabular}{|c|c|c|}
\hline
Parameter & Value & Ref. \\
\hline
$f_{B_d} \sqrt{B_{B_d}}$   & $0.23 \pm 0.04$ GeV & \cite{Buras_2001}\\
$f_D \sqrt{B_D}$           & $\geq 0.2$ GeV & \cite{UKQCD_2000}\\
$\xi=\frac{f_{B_s}\sqrt{B_{B_s}}}{f_{B_d}\sqrt{B_{B_d}}}$& $1.15 \pm 0.06$ & \cite{Buras_2001}\\
$f_K$                      & 0.16 GeV & \cite{Buras_2001}\\
$B_K$                      & $0.85 \pm 0.15$ &\cite{Buras_2001}\cr
\hline
\end{tabular} }
\caption{ \small { Values of decay constants and bag parameters used in the analysis. }}
\label{tab:theoretical_data}
\end{center}
\end{table}

The rare Kaon decays which we consider are $K_L \to \mu \bar
\mu$ and $K^+ \to \pi^+ \nu \bar \nu$.
A bound on $B(K_L \to \mu \bar \mu)_{SD}$ (the short--distance contribution to the
dispersive part of $B(K_L \to \mu \bar \mu)$) can be extracted from
the experimental data as described e.g. in ref. \cite{Buras_2001}.
The expression for this quantity in the four generation model is given by
\begin{equation}
  \label{eq:kmumu}
    B(K_L \to \mu \bar \mu)_{SD}=6.32\times10^{-3} \times {\left [Y_{NL}
        Re\lambda_c^{sd} + {\eta_t}^Y Y_0(x_t) Re\lambda_t^{sd} +
        {\eta_{t^\prime}}^Y Y_0(x_{t^\prime})
        Re\lambda_{t^\prime}^{sd}\right ]}^2,
\end{equation}
where $Y_{NL} = (3.5\pm 0.6)\times 10^{-4}$
\cite{Buras_1996} represents the charm contribution.
In our analysis we neglect the QCD
correction factors ${\eta_t}^Y$ and ${\eta_{t^\prime}}^Y$, since they are
close to unity.
The expression for $B(K^+ \to \pi^+ \nu \bar \nu)$ in the four generation model is given by
\begin{equation}
    B(K^+ \to \pi^+ \nu \bar \nu) =
    1.55\times10^{-4} \times
    \left | \lambda_t^{sd} \eta_t^X X_0(x_t) + \lambda_{t^\prime}^{sd}
      \eta_{t^\prime}^X X(x_{t^\prime}) + X_{NL} \lambda_c^{sd} \right |^2,
\end{equation}
where $X_{NL}=(9.8\pm 1.4)\times 10^{-4}$ represents the
charm contribution.
In this case we again neglect the QCD corrections
which are close to unity. We also neglect the uncertainty in $X_{NL}$.

Next we discuss mass differences in the various neutral meson systems.
The four generation expression for the short--distance contribution to
$\Delta m_K$ is given by
\begin{eqnarray}
  \lefteqn {
    \Delta m_K^{SD} = \frac {G_F^2 m_W^2 f_K^2 B_K m_K}{6 \pi^2}
    Re[{\lambda_c^{sd}}^2 \eta_c^K S_0(x_c) +
    {\lambda_t^{sd}}^2 \eta_t^K S_0(x_t) +
    {\lambda_{t^\prime}^{sd}}^2 \eta_{t^\prime}^K S_0(x_{t^\prime})+{} } \nonumber\\
  \nonumber \\
  & & {} + 2 \lambda_c^{sd} \lambda_t^{sd} \eta_{ct}^K S_0(x_c,x_t) +
  2{\lambda_c^{sd}}{\lambda_{t^\prime}^{sd}} \eta_{ct^\prime}^K S_0(x_c,x_{t^\prime}) +
  2{\lambda_t^{sd}}{\lambda_{t^\prime}^{sd}} \eta_{tt^\prime}^K S_0(x_t,x_{t^\prime})].
\end{eqnarray}
Assuming that the long--distance contributions can be at
most of the order of the experimental value $\Delta m_K^{exp}$,
we take this constraint as
$0 \lesssim \Delta m_K^{SD} \lesssim 7\times 10^{-15}$ GeV.

The contributions to the short--distance part of the $D^0-\bar{D^0}$ meson mixing
amplitude $M_{12}$ in the SM are very small compared to the current
experimental bound (by several orders of magnitude). The leading
diagram for $M_{12}^{SD}$ in the four generation case contains two
$b^\prime$ quarks in the loop, and is given by
\begin{equation}
  \label{eq:delta_md_expr}
  |\ M_{12}^{SD}\ | = \frac {G_F^2 m_W^2 f_D^2 B_D m_D}{12 \pi^2}
  {|{V_{cb^\prime}^*} {V_{ub^\prime}^*}|}^2 \eta_{b^\prime}
  S_0(x_{b^\prime}) \ .
\end{equation}
Since the long--distance contributions to $M_{12}$ are also
estimated to be small compared to the experimental bound, we demand in the
four--generation case
$\left | M_{12}^{SD} \right | \leq 6.3\times 10^{-14}$ GeV,
where the experimental bound was taken as in table
\ref{tab:experimental_data} with $\eta_{b^\prime}=0.56$.
The experimental bound we use is weaker than the one usually considered (e.g. in \cite{PDG_2000}) since
it does not assume $\delta=0$ and uses instead the available
experimental data \cite{RAZ_2001}.
From the parametrization of $V_{CKM}$,
${|V_{ub^\prime} V_{cb^\prime}|}^2 \approx \sin^2\theta_{14}\sin^2\theta_{24}$,
so eq.~\eqref{eq:delta_md_expr} gives the bound
\begin{equation}
  \label{eq:delta_md_bound}
  \sin\theta_{14}\sin\theta_{24} \lesssim
  \left\{\begin{array}{l} 4.1\times10^{-3}\ \ \ \ m_{t^\prime}=500 \ {\rm
  GeV}\\
  9.1\times10^{-3}\ \ \ \ m_{t^\prime}=200 \ {\rm GeV} \end{array} \right.
\end{equation}

The four generation expression for $\Delta m_{B_d}$ is given by
\begin{equation}
  \label{eq:delta_mbd}
  \Delta m_{B_d} = \frac {G_F^2 m_W^2 f_{B_d}^2 B_{B_d} m_{B_d}}{6 \pi^2}
  \left |{\lambda_t^{bd}}^2 \eta_t^B S_0(x_t) +
  {\lambda_{t^\prime}^{bd}}^2 \eta_{t^\prime}^B S_0(x_{t^\prime}) +
  2{\lambda_t^{bd}}{\lambda_{t^\prime}^{bd}} \eta_{tt^\prime}^B
  S_0(x_t,x_{t^\prime}) \right | \ .
  \end{equation}
We take the QCD factors as $\eta_t^B=0.55$,
$\eta_{tt^\prime}^B=0.5$ and $\eta_{t^\prime}^B=0.54$.

The expression for $\Delta m_{B_s}$ is the same as that for $\Delta
m_{B_d}$, the only change is to replace $d \to s$ in all
places. Using the input of tables \ref{tab:experimental_data} and \ref{tab:theoretical_data}, one obtains
\begin{equation}
  \left | \frac {{\lambda_t^{bs}}^2 \eta_t^B S_0(x_t) +
      {\lambda_{t^\prime}^{bs}}^2 \eta_{t^\prime}^B S_0(x_{t^\prime}) +
      2{\lambda_t^{bs}}{\lambda_{t^\prime}^{bs}} \eta_{tt^\prime}^B
      S_0(x_t,x_{t^\prime})}
    {{\lambda_t^{bd}}^2 \eta_t^B S_0(x_t) +
      {\lambda_{t^\prime}^{bd}}^2 \eta_{t^\prime}^B S_0(x_{t^\prime}) +
      2{\lambda_t^{bd}}{\lambda_{t^\prime}^{bd}} \eta_{tt^\prime}^B
      S_0(x_t,x_{t^\prime})} \right | \gtrsim 19.6 \ .
\end {equation}

Next we consider the $\varepsilon_K$ constraint. In a four generation
model, one has
\begin{eqnarray}
\lefteqn {\varepsilon_K = \frac{1}{2} C_\varepsilon B_K \times Im[{\lambda_c^{sd}}^{*^2} \eta_c^K
S_0(x_c)+{\lambda_t^{sd}}^{*^2} \eta_t^K S_0(x_t)+ {\lambda_{t^\prime}^{sd}}^{*^2} \eta_{t^\prime}^K
S_0(x_{t^\prime}) \ + }  \nonumber\\
\nonumber \\
& & {} +2{\lambda_c^{sd}}^*{\lambda_t^{sd}}^* \eta_{ct}^K
S_0(x_c,x_t) + 2{\lambda_c^{sd}}^*{\lambda_{t^\prime}^{sd}}^*
\eta_{ct^\prime}^K S_0(x_c,x_{t^\prime}) +
2{\lambda_t^{sd}}^*{\lambda_{t^\prime}^{sd}}^* \eta_{tt^\prime}^K S_0(x_t,x_{t^\prime})]
\ .
\end{eqnarray}
where $C_\varepsilon \equiv G_F^2 m_W^2 f_K^2 m_K / (6 \pi^2\sqrt{2}
\Delta m_K)=3.8\times 10^4$.
For the SM QCD corrections we take~\cite{Buras_2001}
$\eta_t^K=0.57$, $\eta_c^K=1.38$ and $\eta_{ct}^K=0.47$.
For the new QCD corrections we take~\cite{Hattori_1999}
$\eta^K_{t^{\prime} t^{\prime}} = 0.57$, $\eta^K_{t t^{\prime}} = 0.6$ and
$\eta^K_{c t^{\prime}} = 0.5$.

Another constraint, $a_{\psi K}$, comes from the decays $B\to \psi K$.
In the SM, this quantity measures to an excellent approximation
$\sin2\beta$, where $\beta \equiv
\arg(-\lambda_c^{bd}/\lambda_t^{bd})$ is one of the angles in the
unitarity triangle.
In the presence of New Physics, this is modified \cite{Babar_1998} to
$\sin(2\beta-2\theta_d)$, where $\beta$ is defined as in the SM and
$\theta_d$ is the New Physics phase of the mixing amplitude $M_{12}$.
In the case of four generations, $\theta_d$ is given by
\begin{equation}
  2\theta_d = -\arg \left (\frac {M_{12}} {M_{12}^{SM}} \right )=
  -\arg \left ( 1 + \frac {\eta_{t^\prime}^B
      S_0(x_{t^\prime}) {\lambda_{t^\prime}^{bd}}^2 + 2 \eta_{t
        {t^\prime}}^B S_0(x_t,x_{t^\prime}) \lambda_t^{bd}
      \lambda_{t^\prime}^{bd}}
    {\eta_t^B S_0(x_t) {\lambda_t^{bd}}^2} \right) ,
  \label{eq:theta_d}
\end{equation}
where $M_{12}$ and $M_{12}^{SM}$ are the mixing amplitudes for the $B$ meson system in
the four--generation model and the SM respectively.
For the experimental data we use the world--average of $a_{\psi K}$, given
in table \ref{tab:experimental_data}, at two sigma.
Recently, a new preliminary result regarding the $B\to \psi K$
measurement was published by the BABAR collaboration
\cite{BABAR_2002}. As this is still only a
preliminary result, we did not add it to our analysis. However,
these results are not expected to have a dramatic impact on the
results obtained here.

In the following subsections we discuss in more detail two additional
constraints used in our analysis, those coming from $\Gamma(Z \to b \bar b)$ and from
$B(B\to X_s \ell^+ \ell^-)$.


\subsection{$\Gamma(Z \to b \bar b)$}
\label{sec:gammaz-to-b-bbar}

In this section we follow the derivation and the notations of
ref.~\cite{BERNABEU_1991}, and add to it the contributions of the
fourth generation. The decay rate for $Z\to q \bar{q}$ can be written
as
\begin {multline}
  \Gamma(Z \to q \bar{q}) = \frac{\hat{\alpha}}{16\hat{s}^2_W
    \hat{c}^2_W}m_Z(|a_q|^2+|v_q|^2)(1+\delta_q^{(0)})(1+\delta^q_{QED})\times\\
  \times (1+\delta_{QCD}^q)(1+\delta^q_{\mu})(1+\delta^q_{tQCD})(1+\delta_b) \ .
\end{multline}
We use a caret for quantities given in the $\overline{MS}$ renormalization scheme.
Here $\hat{\alpha}$ is the electromagnetic fine structure constant,
$\hat{s}^2_W$ and $\hat{c}^2_W$ are, respectively, the sine and cosine
squared of the Weinberg angle, $a_q = 2I_3^q$ and $v_q =
2I_3^q-4Q_q \hat{s}^2_W$ are the relevant axial and vector coupling
constants respectively, and the $\delta$ terms are corrections due to
various high order loops. In the calculation of $R_b$, most of the
$\delta$ terms cancel out and are therefore irrelevant for our discussion.
The terms which remain are $\delta_{tQCD}^q$, which consists of QCD
contributions to the axial part of the decay, $\delta_\mu^q$ which
consists of kinematical effects of external fermion masses (including
mass--dependent QCD corrections), and $\delta_b$ which is different
from zero only for $q = b$ and is due to the $Z \to b
\bar{b}$ vertex loop corrections.

The contributions to the corrections $\delta_{tQCD}^q$ originate from doublets with large
mass splitting, and are given in the four--generation case by
\begin{eqnarray}
  \label{eq:delta_QCD}
  \delta_{tQCD}^q =- \frac{a_q}{v_q^2+a_q^2} {\left (
      \frac{\alpha_s}{\pi} \right )}^2 [a_t f(\mu_t) + a_{t^\prime}
  f(\mu_{t^\prime}) + a_{b^\prime} f(\mu_{b^\prime}) ] \ , \nonumber\\
  f(\mu_f) \approx \log \left (\frac{4}{\mu^2_f} \right ) - 3.083+
      \frac{0.346}{\mu^2_f}+\frac{0.211}{\mu_f^4} \, ,
\end{eqnarray}
where $\mu^2_f \equiv 4m^2_f/m^2_Z$.
The corrections to the $Z \to b \bar{b}$ vertex in the four generation
case are given approximately (for $m_{t^\prime}=500$ GeV) by
\begin{equation}
  \label{eq:Zbb_vertex_correction}
  \begin{split}
  \delta_b &\approx 10^{-2}\left [\left (-\frac{m_t^2}{2m_Z^2}+0.2
  \right ) {|V_{tb}|}^2
  + \left (-\frac{m_{t^\prime}^2}{2m_Z^2} + 0.2 \right ){|V_{t^\prime
      b}|}^2 \right]\\
&=
  -0.0154{|V_{tb}|}^2 - 0.148{|V_{t^\prime b}|}^2
\end{split}
\end{equation}
The corrections $\delta_\mu^q$ are given in ref.~\cite{BERNABEU_1991},
and are not affected by the fourth generation.
Using the expressions given in ref.~\cite{BERNABEU_1991} for
$R_s, R_c, R_u$,
together with eqs.~\eqref{eq:delta_QCD} and \eqref{eq:Zbb_vertex_correction}, one obtains
\begin{equation}
  \label{eq:vtb_vtbprime_zbb}
  R_b = (1+2/R_s+1/R_c+1/R_u)^{-1} =
    {\left [ 1 + \frac{3.5837}{1-0.0154{|V_{tb}|}^2 -
        0.148{|V_{t^\prime b}|}^2} \right ]}^{-1}\ .
\end{equation}
Eq.~\eqref{eq:vtb_vtbprime_zbb} together with the experimental
bound taken from table~\ref{tab:experimental_data} at
95$\%$ CL imply ${|V_{tb}|}^2 + 9.5{|V_{t^\prime b}|}^2 \leq 1.14$.
This can be used, together with the unitarity of the third column of
$V_{CKM}$ and the parametrization of $V_{CKM}$, to get
\begin{equation}
  \label{eq:mixing_bounds2}
  \sin\theta_{34}\lesssim
  \left\{\begin{array}{l} 0.14\ \ \ \ m_{t^\prime}=500 \ {\rm
  GeV}\\
  0.3\ \ \ \ m_{t^\prime}=200 \ {\rm GeV} \end{array} \right.
\end{equation}

This bound becomes weaker as $m_{t^\prime}$ is lowered, and is only
very weakly dependent on $m_{b^\prime}$.
In the degenerate case, $m_t=m_{t^\prime}$, one cannot put any limitation on
the mixing angle $\theta_{34}$, as expected.

This constraint contributes significantly to the analysis.
It excludes the maximal mixing solution found in \cite{Hattori_1999}, as
well as some of the other mixing solutions suggested there.
In addition, it probably implies that the up quark mass matrix has
some special structure. This can be seen if we recall the assumption
$m_t \lesssim m_{t^\prime} \lesssim 500$ GeV, which
means that there is no hierarchy between the third and the
fourth generation up quark masses.
One then naively expects large mixing between the
third and the fourth generation, so that $\sin\theta_{34} = O(1)$.
However, the bound on $\sin\theta_{34}$
(eq. \eqref{eq:mixing_bounds2})
implies that in the four--generation model this naive expectation
is not fulfilled, and the mass matrix has some non trivial structure.


\subsection{$B(B\to X_s \ell^+ \ell^-)$}
\label{sec:B_to_Xll}

In previous analyzes (e.g. \cite{Hattori_1999}), only the leading
contributions of Z-mediated diagrams to this branching ratio were
considered. However in light of the improvement in the
experimental data, we use a more detailed analysis of this
quantity. In this section we use mainly the derivation and the
notations of \cite{Buras_1996}. In our analysis, however, we
completely neglect QCD corrections. The decay can be described at
low energies by using the following effective Lagrangian :
\begin{multline}
  \mathcal{L}_{eff}=-\frac{G_F}{\sqrt{2}}\lambda_t^{sb} \left [ C_2
    {(\bar{s}c)}_{V-A}{(\bar{c}b)}_{V-A} + C_{7\gamma}
    \frac{e}{8\pi^2} m_b \bar{s} \sigma^{\mu \nu}(1+\gamma_5)b F_{\mu
      \nu}+ \right.\\
    \left. + C_{9V} {(\bar{s}b)}_{V-A}{(\bar{l}l)}_{V}+
    C_{10A} {(\bar{s}b)}_{V-A}{(\bar{l}l)}_{A} \right ] \ ,
\label{eq:effective_lagrangian_bsll}
\end{multline}
where $(q_1 q_2)_{V \pm A} \equiv q_1 \gamma_\mu(1 \pm \gamma_5)
q_2 $ and $m_b$ is the $b$ quark mass.
The coefficients $C_{7\gamma}, C_{9V}$ and $C_{10A}$ originate
in the SM from electroweak magnetic penguin diagrams, Z and
$\gamma$ penguin diagrams and W box diagrams. The coefficient $C_2$
contributes to the decay via an intermediate $c \bar c$ loop.
In order to get the total branching ratio, one should integrate the differential
branching ratio given in \cite{Buras_1996}
over the entire allowed range, which we take as $\frac{4m_l^2}{m_b^2}
\leq s \leq 1$ (here $s\equiv q^2_{\ell^+ \ell^-}/m_b^2$, where $q$ is the total dilepton momentum).
The resulting bound (for electrons in the final state), when neglecting
QCD corrections, is
\begin{multline}
  \frac{|\lambda_t^{bs}|^2}{|V_{cb}|^2} \left \{
  (0.5|\tilde{C}_9|^2 + 0.5|\tilde{C}_{10}|^2 + 0.93|C_2|^2 + 1.1
  Re[C_2^* \tilde{C}_9] + 125.6 |C_{7\gamma}|^2 + \right.\\
  \left. +4 Re \left [C_{7\gamma}^*
  \tilde{C}_9 \right ] + 4.3 Re [C_2 C_{7\gamma}^*] \right \} \lesssim 42.4,
\label{eq:full_br_ratio_bsll}
\end{multline}
where $\tilde{C}_9 \equiv \frac{2\pi}{\alpha} C_{9V}$
and $\tilde{C}_{10} \equiv \frac{2\pi}{\alpha} C_{10A}$.

The values of the coefficients $C_2, C_{7\gamma}, \tilde{C}_{9}$ and
$\tilde{C}_{10}$ are given in the SM by
\begin{align}
  \begin{split}
    C_2 &= 1,\\
    C_{7\gamma} &= -\frac{1}{2}D_0^\prime(x_t),\\
    \tilde{C}_{9} &=
    \frac{8}{9} \ln \left (
  \frac{M_W}{m_B} \right ) + \frac{Y_0(x_t)}{\sin^2\theta_W}-4Z_0(x_t) ,\\
    \tilde{C}_{10} &=
    - \frac{Y_0(x_t)}{\sin^2\theta_W},
  \end{split}
\end{align}
where $D_0^\prime(x_t), Y_0(x_t), Z_0(x_t)$ are Inami--Lim functions.
When considering a model with a fourth generation, the coefficients
$C_2$, $C_{7\gamma}$, $\tilde{C}_9$ and $\tilde{C}_{10}$ change as follows:
\begin{align}
  \begin{split}
  C_2 &= 1+\frac{\lambda_{t^\prime}^{bs}}{\lambda_t^{bs}},\\
  C_{7\gamma}(M_W) &= -\frac{1}{2} \left (D_0^\prime(x_t) + \frac{\lambda_{t^\prime}^{bs}}{\lambda_t^{bs}}D_0^\prime(x_{t^\prime})
  \right ),\\
  \tilde{C}_9 &= \frac{8}{9} \ln \left (\frac{M_W}{m_B}
  \right ) \left ( 1+\frac{\lambda_{t^\prime}^{bs}}{\lambda_t^{bs}}
  \right )
  + \frac{Y_0(x_t) + \frac{\lambda_{t^\prime}^{bs}}{\lambda_t^{bs}}Y_0(x_{t^\prime})}{\sin^2\theta_W} -
  4 \left [Z_0(x_t)+ \frac{\lambda_{t^\prime}^{bs}}{\lambda_t^{bs}}Z_0(x_{t^\prime}) \right],\\
  \tilde{C}_{10} &= -\frac{Y_0(x_t) + \frac{\lambda_{t^\prime}^{bs}}{\lambda_t^{bs}}Y_0(x_{t^\prime})}{\sin^2\theta_W}.
\end{split}
\label{eq:ABC_Bsll_4thgen}
\end{align}
We use in our analysis the bound of eq. \eqref{eq:full_br_ratio_bsll},
with the coefficients as in eq. \eqref{eq:ABC_Bsll_4thgen}.


\subsection{Numerical results}
\label{sec:numerical-results}

In the numerical analysis we take all the constraints that were
described in the previous subsections, and scan over the allowed
region in the parameter space (namely, the mixing angles according
to bounds given in eq.~(\ref{eq:mixing_bounds}) and
(\ref{eq:mixing_bounds2}) and the three phases between 0 and
$2\pi$). The scan is performed by randomly choosing the values of
the mixing parameters (using a uniform distribution), and checking
for each such point in the parameter space that all the
constraints are met at two sigma. When extracting from the scan
the allowed ranges for various quantities (such as mixing angles
and predictions for observables) we used only 95$\%$ of the data
points given by the scan results. In this section we analyze the
results for the case of $m_{t^\prime}=500$ GeV. The results for
$m_{t^\prime}=200$ GeV are given in the following section.

\begin{figure}[htb]
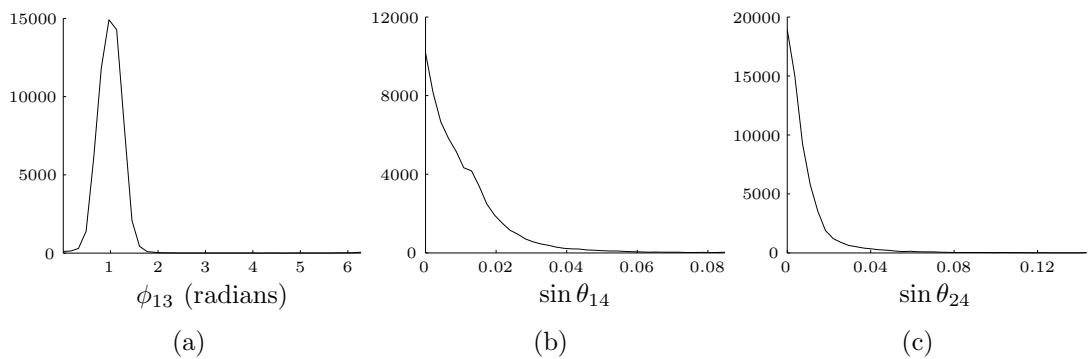

  \begin{center}
    \subfigure[]
    {
      \label{fig:phi13_hist}
      \input{phi13.tex}
    }
    \subfigure[]
    {
      \label{fig:theta14_hist}
      \input{theta14.tex}
    }
    \subfigure[]
    {
      \label{fig:theta24_hist}
      \input{theta24.tex}
    }
    \caption{Histograms of (a) $\phi_{13}$, (b) $\sin\theta_{14}$ and
    (c) $\sin\theta_{24}$ for 60,000 data points, for $m_{t^\prime}=500$
    GeV and $m_{b^\prime}=470$ GeV.}
    \label{fig:histograms}
  \end{center}
\end{figure}

According to the scan, some of the new mixing angles and phases have
restricted ranges (see figure \ref{fig:histograms}):
\begin{align}
  \label{eq:mixing_params_numerical_bounds}
  0 &\lesssim \phi_{13} \lesssim \frac{\pi}{2}, \nonumber\\
  0 &\lesssim \sin\theta_{14} \lesssim 0.031\ , \nonumber\\
  0 &\lesssim \sin\theta_{24} \lesssim 0.033\ , \\
  0 &\lesssim \sin\theta_{34} \lesssim 0.14\ . \nonumber
\end{align}
The remaining new phases ($\phi_{14}$ and $\phi_{24}$) can take the entire scanned range.

Note that in the SM the allowed range (at two sigma) for
$\phi_{13}$ is $\pi/5 \lesssim \phi_{13} \lesssim 4\pi /9$. This
is quite close to the range of this phase in the four generation
model. But while in the SM the allowed range is dominated by the
$\varepsilon_K$ constraint together with $\Delta m_{B_d}$ and
$\Delta m_{B_s}$, in the four--generation model it is influenced
by a correlated effect of several constraints (and not only the
three that dominate in the SM case).

\begin{figure}[htb]
  \begin{center}
    \subfigure[]
    {
      \label{fig:NP_params_sd}
      \resizebox{4cm}{4cm}{\includegraphics{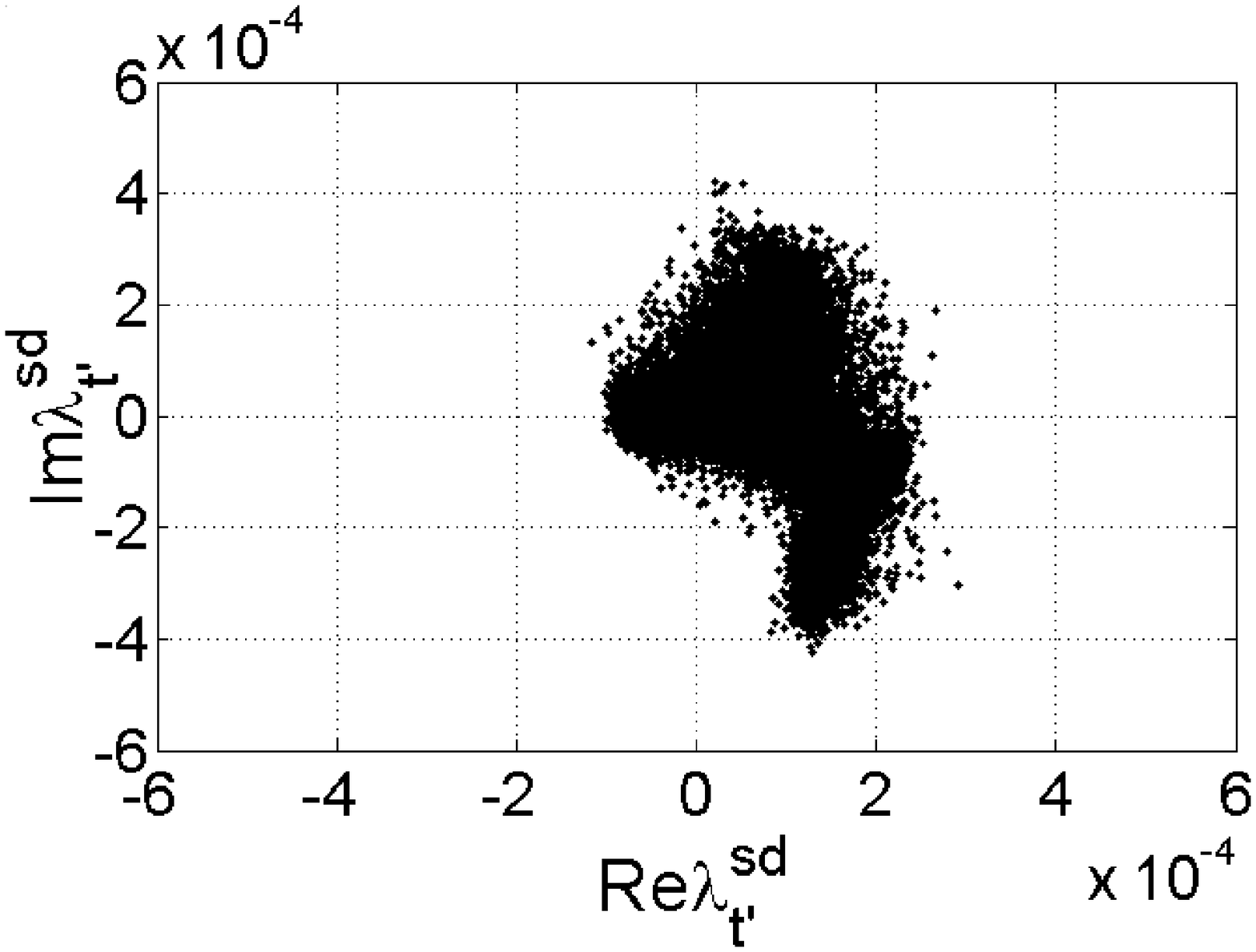}}
    }
    \subfigure[]
    {
      \label{fig:NP_params_bd}
      \resizebox{4cm}{4cm}{\includegraphics{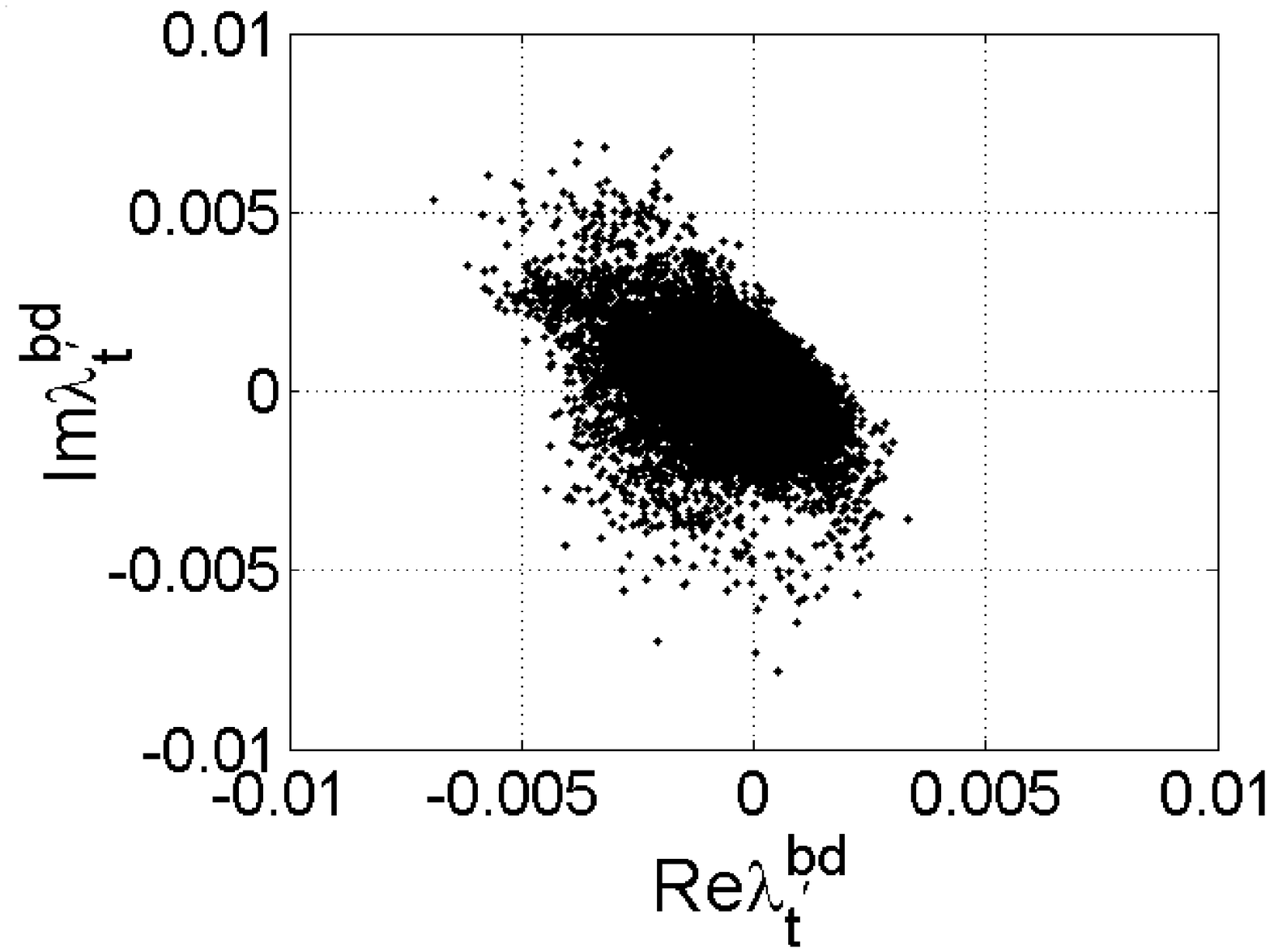}}
    }
    \subfigure[]
    {
      \label{fig:NP_params_bs}
      \resizebox{4cm}{4cm}{\includegraphics{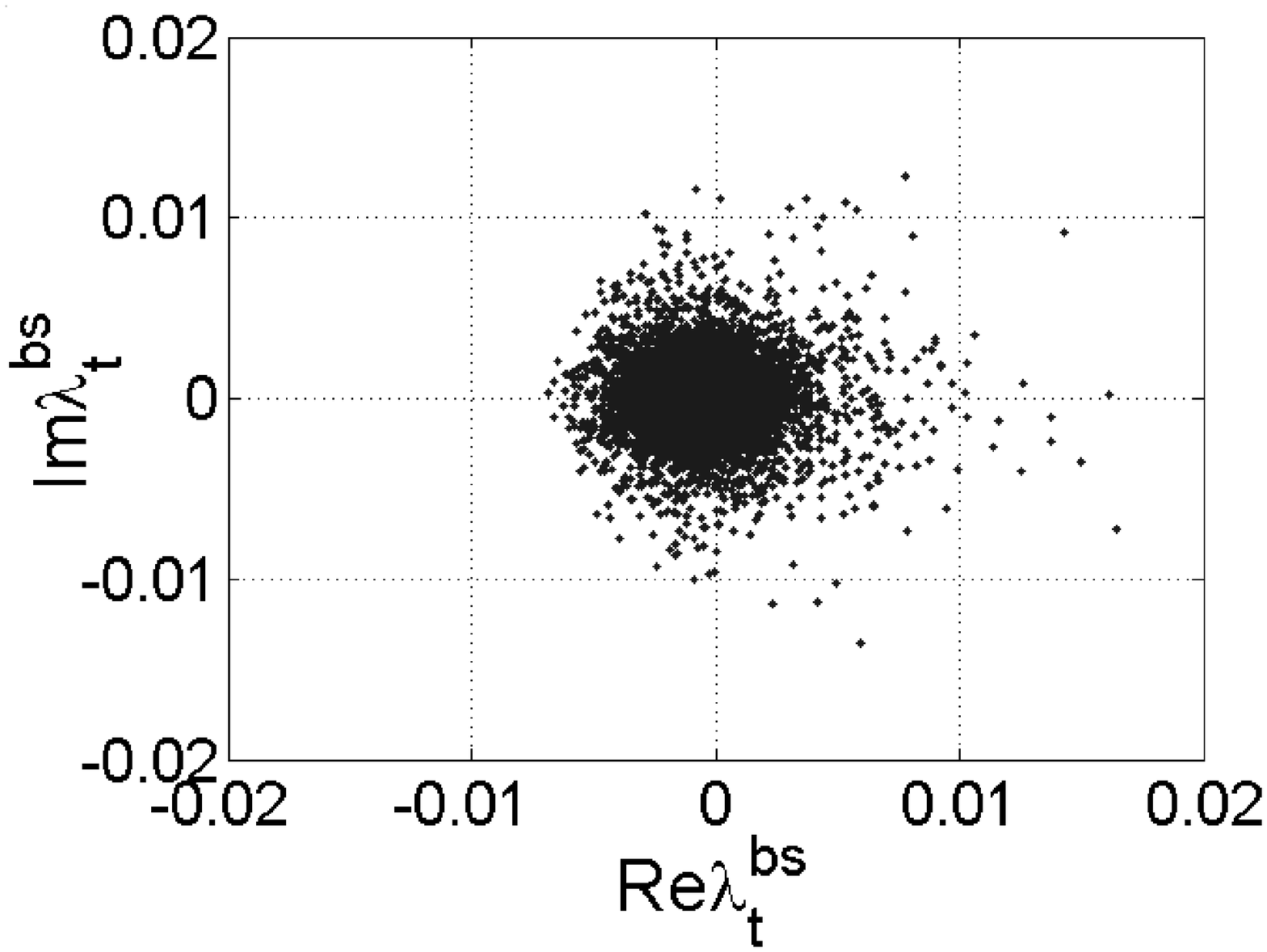}}
    }
    \caption{Scatter plot of the New Physics parameters
    (a) $\lambda_{t^\prime}^{sd}$, (b) $\lambda_{t^\prime}^{bd}$ and
        (c) $\lambda_{t^
        \prime}^{bs}$ with all the constraints, for $m_{t^\prime}=500$
        GeV and $m_{b^\prime}=470$ GeV.}
    \label{fig:NP_params}
  \end{center}
\end{figure}

In order to see in a clearer way the impact of the various
constraints on New Physics, one can look at the allowed regions of the parameters
$\lambda_{t^ \prime}^{sd}$, $\lambda_{t^ \prime}^{bd}$ and $\lambda_{t^ \prime}^{bs}$.
Their scatter plots are given in Figure~\ref{fig:NP_params}.
Our analysis shows that there is very little correlation between the
various constraints. This means that each of the parameters is
effectively constrained only by a small number of constraints.
We now consider each of the parameters and the constraints that
influence it most.
\begin{enumerate}
\item
  {
      $\lambda_{t^ \prime}^{sd}$: The bound on
      $Im\lambda_{t^ \prime}^{sd}$ comes mainly from the $B(K^+ \to \pi^+
      \nu \bar \nu)$ constraint.
      The real part, $Re\lambda_{t^\prime}^{sd}$, is mainly constrained by
      $B(K_L \to \mu \bar \mu)_{SD}$.
      The general shape of $\lambda_{t^\prime}^{sd}$ is mainly due to the
      $\varepsilon_K$ constraint.
    }
  \item
    {
      $\lambda_{t^\prime}^{bd}$: $|\lambda_{t^\prime}^{bd}|$ is mainly constrained by the $\Delta
      m_{B_d}$ constraint. The $a_{\psi K}$ constraint is responsible for the area excluded in
      the upper--right and lower--left regions in figure \ref{fig:NP_params_bd}.
      We also examined how an improvement in the experimental error for
      $a_{\psi K}$ will affect the influence of this constraint.
      As it turns out, the excluded area is almost insensitive to the size
      of the error, and even an error as low as 0.01 (10 percent of the current error) will
      have very little effect in the $(Re\lambda_{t^\prime}^{bd},
      Im\lambda_{t^\prime}^{bd})$ plane.
      A change in the central value from 0.8 to 0.7 will have
      some influence on the excluded area. In general, the higher the
      central value is, the larger is the excluded area. However, the
      modifications in the excluded area are in any case not drastic.
      This means that the coming experimental results for
      $a_{\psi K}$ are not expected to change the bounds on the
      fourth--generation flavor parameters in a significant manner.
    }
    \item
      {
        $\lambda_{t^\prime}^{bs}$: This parameter is constrained
        mainly by the $B\to X_s \ell^+ \ell^-$ bound, together with the tree--level
        constraints and the $Z\to b \bar{b}$ constraint. It is little
        influenced by the $\Delta m_{B_s}$ bound.
        Our bounds for this parameter are better than those previously
        quoted in the literature, since we use an updated (and
        stronger) experimental constraint for $B\to X_s \ell^+ \ell^-$.
      }
    \end{enumerate}

A good way to compare the four--generation model to the SM is to
look at the predictions given by the two models for various
quantities.
The results are summarized in table \ref{tab:predictions1}.
\begin{table}[htb]
\begin{center}
\footnotesize{
\begin{tabular}{|c|c|c|c|c|}
\hline
 & $B(K_L\to \mu \bar \mu)_{SD}$ & $B(K_L\to \pi^0 \nu \bar \nu)$ &
$B(K^+\to \pi^+ \nu \bar \nu)$ & $B(B\to X_s \ell^+ \ell^-)$\\
\hline
\hline
 & & & &\\
experiment         &
$\leq 3.75\times 10^{-9}$  &
$\leq 5.9\times 10^{-7}$ &
$\leq 5.07\times 10^{-10}$ &
$\leq 1 \times 10^{-5}$\\ & & & &\\
SM                 & $[0.4 , 1.4]\times 10^{-9}$ & $[1.7 ,
5.6]\times 10^{-11}$ &
$[4.7 , 11.0]\times 10^{-11}$ &
$[3.8, 13] \times 10^{-6}$ (ref. \cite{ALI_1997})\\ & & & &\\
four--generations &
$\leq 1.18\times 10^{-8}$  &
$\leq 4.2\times 10^{-9}$ &
$\leq 6.7\times 10^{-10}$ & - \\ & & & &\\
VDQ & $\leq 2.4\times 10^{-8}$  & $\leq 4\times 10^{-10}$
& $\leq 8.3\times 10^{-10}$ & $\leq 2.5 \times 10^{-3}$\\ & & & &\\
\hline
\hline
 & $\Delta m_{B_s}\, [ps^{-1}]$ & $|M_{12}^D|\,$[GeV] &
$\frac{a_{SL}}{{\left (\Gamma_{12}/M_{12} \right )}^{SM}}$ & \\
\hline
\hline
& & & &\\
experiment         & $\geq 15$  & $\leq 8.2\times 10^{-14}$  &
$[-5.84 , +5.05]$ \cite{ALEPH_2001,BABAR_2001,CLEO_2001,OPAL_2000} &
 \\ & & & &\\
SM                 & $[15 , 32]$ & $\sim 10^{-17}$ to $10^{-16}$
& $[0.04, 0.26]$ (Ref.~\cite{LAPLACE_LIGETI_NIR_PEREZ_2002}) &
 \\ & & & &\\
four--generations &
$[12 , 28]$  &
$\leq 2.7\times 10^{-15}$ &
$[-1.3 , +1.8]$ &
 \\ & & & &\\
VDQ &
$[15 , 32]$  &
- &
$[-0.47 , +0.28]$ &
 \\ & & & &\\
\hline
\end{tabular} }
\caption{ \small { Predictions for different quantities in the SM and
 in the four--generation model compared to the experimental
 bounds, taken at two sigma. The category of 'favored range' in the four--generation
 model refers to the range that contains 95$\%$ of the data points.}}
\label{tab:predictions1}
\end{center}
\end{table}
We give here several notes regarding these predictions:
\begin{enumerate}
  \item
    {
      The predictions for $|M_{12}^D|$, $B(K_L\to \mu \bar \mu)_{SD}$,
      $B(K^+\to\pi^+ \nu \bar{\nu})$ and $B(K_L\to\pi^0 \nu \bar{\nu})$
      in the four--generation model can be significantly higher then
      the SM predictions.
      In the cases of $B(K_L\to \mu \bar \mu)_{SD}$ and
      $B(K^+\to\pi^+ \nu \bar{\nu})$ they can span the entire range
      implied from the experimental constraints.
    }
  \item
    {
      The CP asymmetry in semi--leptonic decays is approximately given by
      the model--independent expression~\cite{LAPLACE_LIGETI_NIR_PEREZ_2002}
      \begin{equation}
        \frac{a_{SL}}{{\left (\Gamma_{12}/M_{12} \right
            )}^{SM}} =
        \frac {\sin 2\theta_d } {
          |M_{12}/M_{12}^{SM} | }
      \end{equation}
      The experimental bound on this quantity is obtained by
      calculating the world--average of $a_{SL}=(0.2\pm 1.4)\times 10^{-2}$ from
      \cite{ALEPH_2001,BABAR_2001,CLEO_2001,OPAL_2000}, and taking
      into account the theoretical predictions for
      ${\left (\Gamma_{12}/M_{12} \right )}^{SM} \approx -(0.79 \pm 0.27)\times 10^{-2}$
      (by an updated scan according to the data in ref. \cite{LAPLACE_LIGETI_NIR_PEREZ_2002}).
      Although our scan improves the bounds which were obtained in
      \cite{Eyal_1999}, the four--generation
      model prediction for this quantity can still be about an order
      of magnitude higher than that of the SM. Also, this quantity in the four
      generation model may have a different sign from that predicted by the SM.
      The four--generation prediction for $a_{SL}$ is quite far from the
      current experimental bound. In case the experimental bound improves
      by about a factor of 3 for the upper bound and 4.5 for the lower bound,
      this bound will become significant in the analysis.
    }
  \item
    {
      Taking all constraints at two sigma, the $a_{\psi K}$ prediction in
      the four--generation model covers the entire range allowed by the $B\to \psi K$ constraint.
      The SM prediction covers the range $0.6 \leq a_{\psi K} \leq
      0.95$. However, the difference between the SM and the four--generation model regarding this
      quantity can be clearly seen if we perform the scan without including the $B\to \psi
      K$ constraint. Then the four--generation model allows the range $-1 \leq a_{\psi K}
      \leq 1$, while the SM allowed range is only $0.4 \leq a_{\psi K} \leq 0.95$.
      Still, from the latest experimental results it is clear
      that the value of $a_{\psi K}$ lies at the higher
      part of the allowed range, so this information is of limited impact.
    }
  \item
    {
      The $\Delta m_{B_s}$ prediction in the four--generation model is
      similar to the prediction of the SM.
      This means that detection of
      $\Delta m_{B_s}$ outside the SM range will be a major problem
      not only for the SM but also for the four generation model.
    }
  \end{enumerate}


\subsection{The case of $m_{t^\prime}=200$ GeV, $m_{b^\prime}=170$ GeV}
\label{sec:mtp_200}

In this case, the bounds that are obtained for the mixing
parameters are significantly weaker than for the case of
$m_{t^\prime}=500$ GeV, $m_{b^\prime}=470$ GeV. The fact that the
strongest effects are obtained for the heavier mass of $t^\prime$
can be explained as follows. Roughly speaking, the diagrams that
we consider put bounds on terms of the form
$\lambda_{t^\prime}^{lm}y_i(x_{t^\prime})$, where
$y_i(x_{t^\prime})$ is some Inami--Lim function. Since
$y_i(x_{t^\prime})$ grows with $x_{t^\prime} \equiv
{m_{t^\prime}}^2/{m_W}^2$, the heavier the mass is, the stronger
is the bound on the parameter $\lambda_{t^\prime}^{lm}$.

The histograms for the phase $\phi_{13}$ and the angles $\sin\theta_{14}$ and
$\sin\theta_{24}$ are qualitatively the same as for the case
$m_{t^\prime}=500$.
The resulting ranges of the new mixing angles and phases are given by:
\begin{align}
  \label{eq:mixing_params_numerical_bounds_m200}
  0 &\lesssim \phi_{13} \lesssim \frac{\pi}{2}, \nonumber\\
  0 &\lesssim \sin\theta_{14} \lesssim 0.051\ , \nonumber\\
  0 &\lesssim \sin\theta_{24} \lesssim 0.056\ , \\
  0 &\lesssim \sin\theta_{34} \lesssim 0.3\ . \nonumber
\end{align}
The remaining new phases ($\phi_{14}$ and $\phi_{24}$) can again take the entire scanned range.

\begin{figure}[htb]
  \begin{center}
    \subfigure[]
    {
      \label{fig:NP_params_bd_m200}
      \resizebox{4cm}{4cm}{\includegraphics{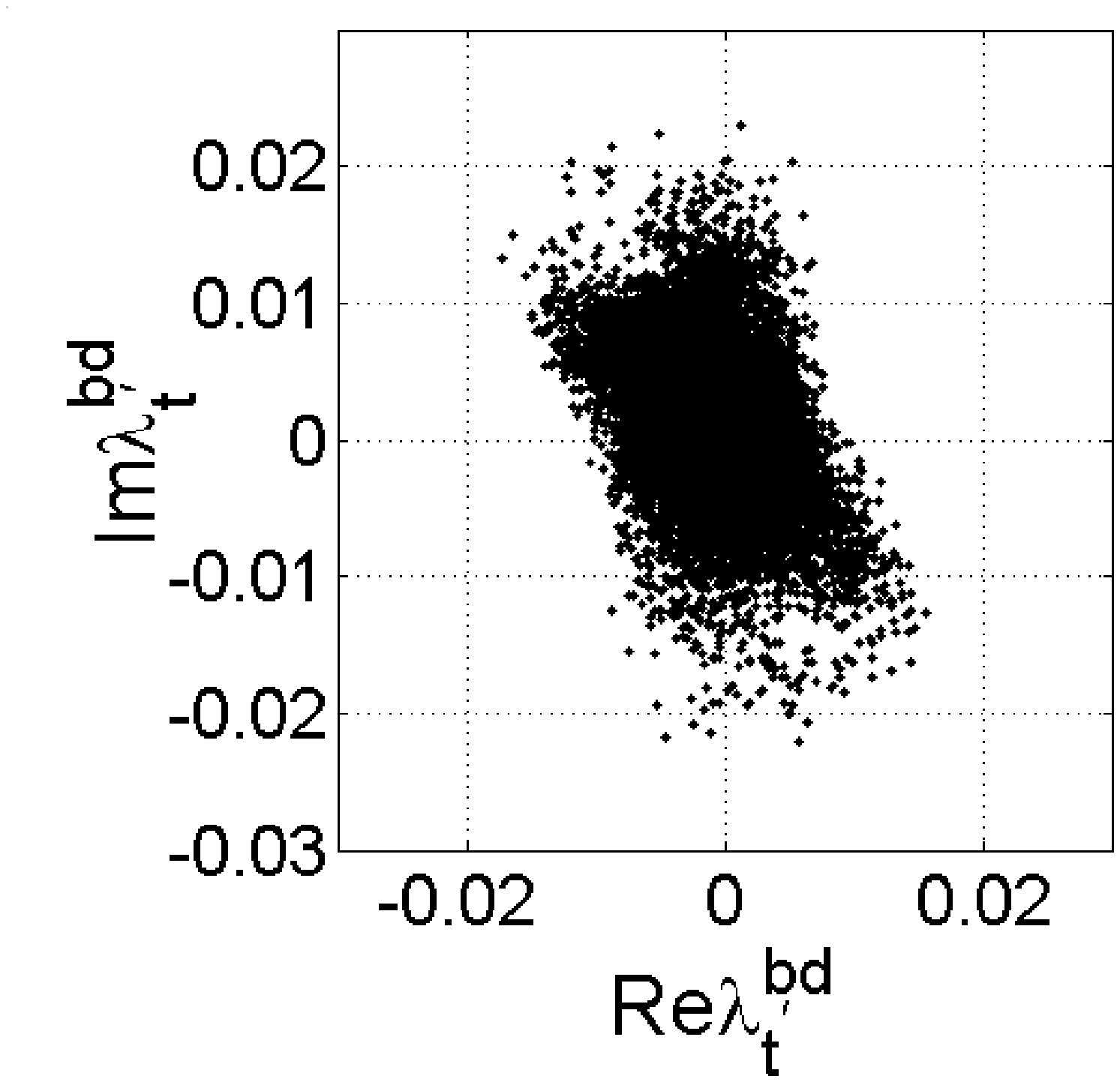}}
    }
    \subfigure[]
    {
      \label{fig:NP_params_bs_m200}
      \resizebox{4cm}{4cm}{\includegraphics{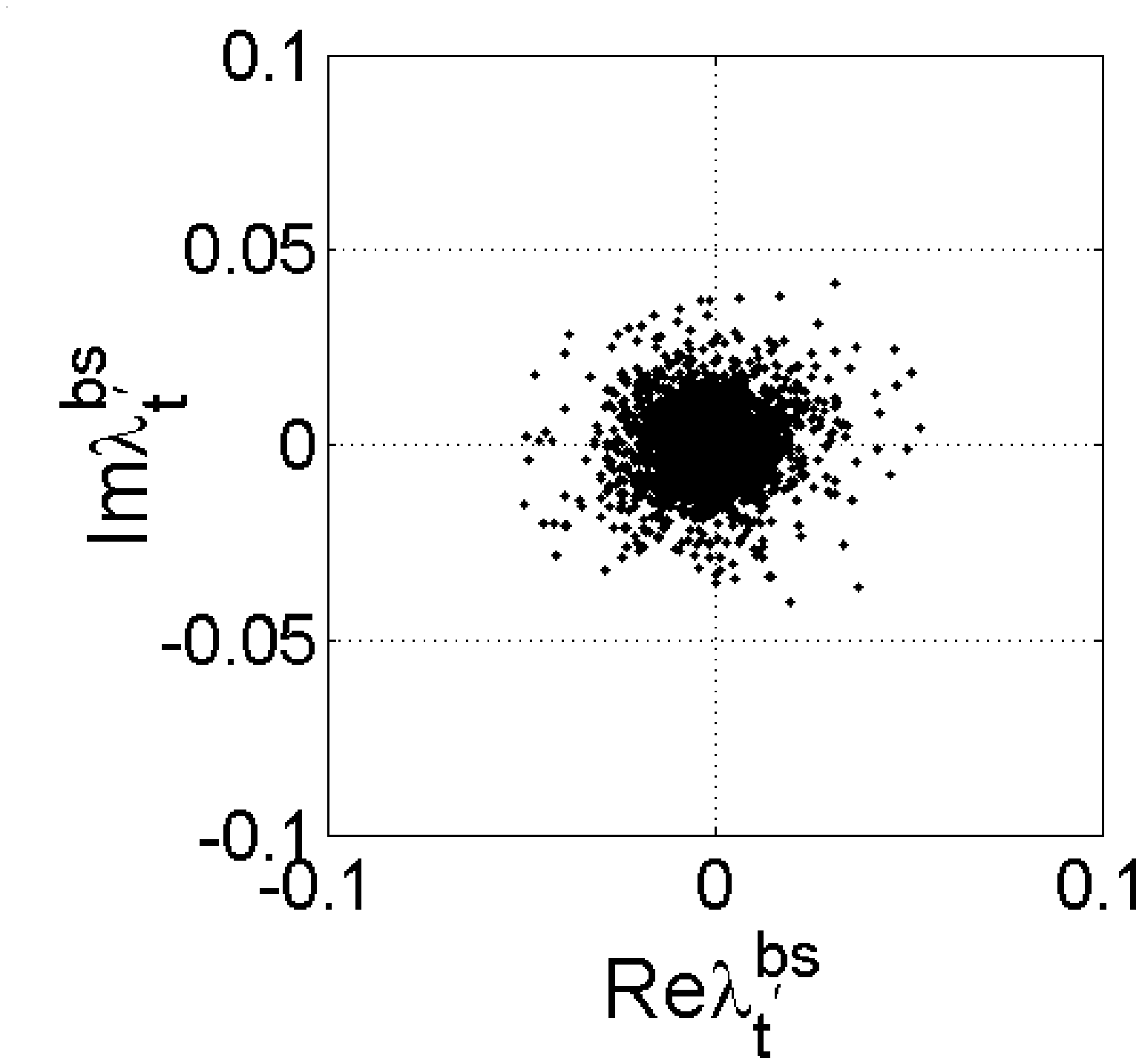}}
    }
    \subfigure[]
    {
      \label{fig:NP_params_sd_m200}
      \resizebox{4cm}{4cm}{\includegraphics{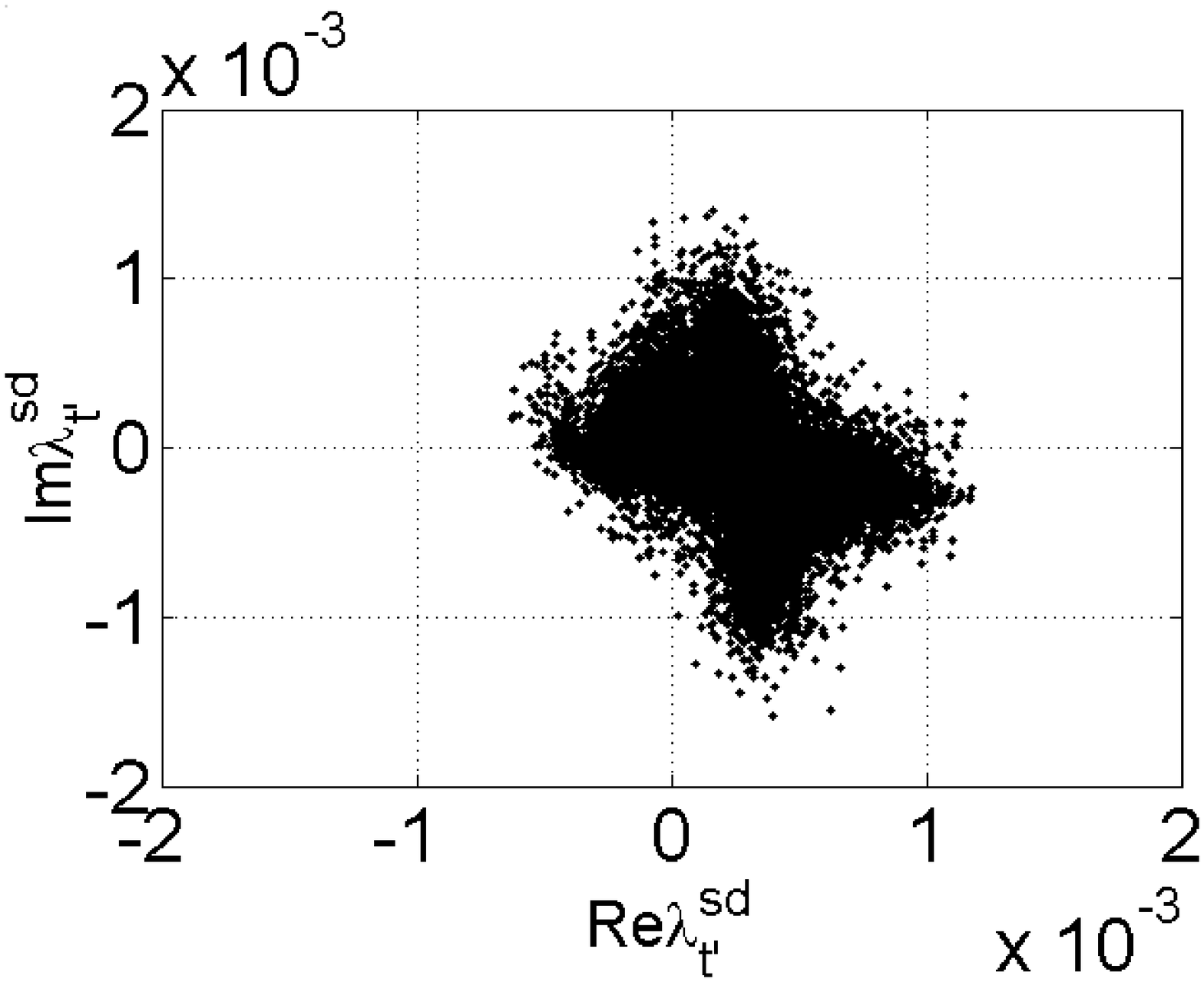}}
    }
    \caption{Scatter plot of the New Physics parameters (a) $\lambda_{t^
        \prime}^{bd}$, (b) $\lambda_{t^ \prime}^{bs}$ and (c) $\lambda_{t^
        \prime}^{sd}$ with all the constraints, for $m_{t^\prime}=200$
        GeV and $m_{b^\prime}=170$ GeV.}
    \label{fig:NP_params_m200}
  \end{center}
\end{figure}

The bounds on the New Physics parameters $\lambda_{t^ \prime}^{bd}$, $\lambda_{t^
  \prime}^{bs}$ and $\lambda_{t^ \prime}^{sd}$ are presented in figure
\ref{fig:NP_params_m200}. In this case, $\lambda_{t^ \prime}^{bs}$ is
constrained mainly by the tree--level decay processes, $Z\to b \bar
b$ and unitarity, and not by $B\to X_s \ell^+ \ell^-$ as in the case of
$m_{t^\prime}=500$ GeV. Other than that, one can see that the bounds
for the cases of $m_{t^\prime}=500$ GeV and $m_{t^\prime}=200$ GeV are
qualitatively the same, with obvious numerical differences.

The predictions for various quantities in the case of $m_{t^\prime}=200$
GeV, $m_{b^\prime}=170$ GeV are quite close to the those in the case of $m_{t^\prime}=500$
GeV, $m_{b^\prime}=470$ (table \ref{tab:predictions1}).
Basically this means that the allowed ranges of the four--generation
model for these quantities is almost independent of the new quark
masses.

\newpage

\section{Vector--like Down--type Quarks (VDQ)}
\label{sec:vldt}

\subsection{Background}
\label{sec:vldt_Introduction}

We consider a model in which a single VDQ is added to the SM. Vector--like quarks
transform as triplets under the ${SU(3)}_C$ symmetry, and both
their left--handed and right--handed components transform as
singlets under the ${SU(2)}_L$ symmetry.
VDQs are predicted by various extensions of the SM, such as grand
unified theories based on the $E_6$ Lie algebra. Also, as explained in
section \ref{sec:introduction}, models with VDQ exhibit interesting
features such as the violation of CKM unitarity and
the related appearance of FCNC contributions at tree--level, and
the appearance of additional CP violating phases.

The constraints we consider in our analysis are the following:
charged--current tree--level decays, the branching ratios $B(B \to
X_s \ell^+ \ell^-), B(K^+\to \pi^+ \nu \bar \nu)$ and
$B(K_{L}\to\mu\bar{\mu})_{SD}$, the mass differences $\Delta
m_{B_d}$ and $\Delta m_{B_s}$, the CP violating parameters
$\varepsilon_K$, $\frac{\varepsilon\prime}{\varepsilon}$ and
$a_{\psi K}$, and precision electroweak measurements related
to the $Z\bar{b} b$ coupling. The scan procedure that we use is similar to that
used for the four generation model. 

This model was previously studied in the literature (see
\cite{VLDT_PREV} and references therein). Regarding $R_b$ and $A^b_{FB}$ (the
forward--backward asymmetry of the $b$ quark) in this model, see
\cite{Tait_2002}. Recent works similar to
the one presented here were performed by
\cite{Barenboim_2001}. We update the experimental bounds used
there, taking into account the new results from Belle and BABAR
collaborations regarding the $B\to\psi K$ measurements. We
also present the predictions of this model for various observables
and compare them to the predictions of the SM.
While this paper was in final stages of writing, another article
\cite{Silverman_2002} was published on this subject.


\subsection{The model}
\label{sec:vldt_the-model}

We consider a single VDQ, denoted $b^\prime$,
added to the SM. This means that the mass matrix in the down sector is
now $4 \times 4$, and it is diagonalized by a $4 \times 4$ matrix $V$.
The down sector charged--current interactions depend on the $3 \times 4$ upper
submatrix of $V$, which plays the role of the CKM matrix, but are
otherwise unchanged.
The neutral--current interactions are given in this model by
 \begin{equation}
   \mathcal{L}_{NC} = {g\over 2\cos{\theta_W}}Z^\mu (\overline{\mathcal{U}^M_L} \gamma_\mu
   \mathcal{U}^M_L - \overline{\mathcal{D}^{M\alpha}_{L}} \gamma_\mu U_{\alpha\beta} \mathcal{D}^{M\beta}_{L}
   -2\sin^2{\theta_W} J^{EM}_\mu)\ ,
   \label{eq:vldt_neutral_current_lagrangian_mass_basis2}
 \end{equation}
 where a sum over repeated indices is implied ($\alpha,\beta=d,s,b$), and we used the definitions
 \begin{equation}
   \mathcal{U} \equiv
   \left(\begin{array}{l} u\\ c\\ t\\ \end{array}\right),
   \ \
   \mathcal{D} \equiv
   \left(\begin{array}{l} d\\ s\\ b\\ \end{array}\right),
   \ \
   U_{\alpha\beta}\equiv \sum_{i=1}^3 V^*_{i \alpha} V_{i \beta} =
   \delta_{\alpha\beta} - V^*_{4\alpha} V_{4\beta}\ .
 \end{equation}
 In eq. \eqref{eq:vldt_neutral_current_lagrangian_mass_basis2},
 $J^{EM}_\mu$ is the electromagnetic current, which contains also
 terms with the new quark $b^\prime$. Note that
 eq. \eqref{eq:vldt_neutral_current_lagrangian_mass_basis2} contains FCNC.
 In all the processes we consider, the leading New Physics contributions
 come from the tree--level FCNC that appear in eq.
 \eqref{eq:vldt_neutral_current_lagrangian_mass_basis2} through the
 quantities $U_{\alpha\beta}$.
 These New Physics contributions usually compete with contributions
 coming from SM loop processes.
 Other New Physics effects, due to $b^\prime$ quarks in loop diagrams, are naturally highly
 suppressed compared to the tree--level contributions, and can be
 safely neglected.
 We can thus view this model at low energies as having the same
 particle content and interactions as the SM (completely ignoring
 the extra $b^\prime$ quark), but having a non unitary CKM matrix.

 Note that the quantities $U_{bs}$, $U_{bd}$ and $U_{sd}$ play two
 important roles. First, they indicate the amount by which the
 three--generation CKM matrix deviates from unitarity.
 Second, they represent the strength of the new contributions to FCNC.
 These quantities play the same role as $-\lambda_{t^\prime}^{ij}$ in the
 four--generation model.

 The matrix $V$ is not a general unitary matrix; some of the phases
 in it can be removed by change of basis. It can be parameterized
 by nine parameters: six mixing angles and three phases. All these
 parameters appear also in the upper $3\times4$ submatrix.
 As in the four--generation model, we use the specific parameterization of
 \cite{Botella_1986} for the $4\times4$ matrix $V$, in order to
 incorporate all the correlations in our analysis. The mixing
 angles are again referred to as $\theta_{12}, \theta_{13},
 \theta_{23}, \theta_{14}, \theta_{24}$ and $\theta_{34}$, and the
 phases as $\phi_{13}, \phi_{14}$ and $\phi_{24}$.


\subsection{The constraints}
\label{sec:vldt_tree-level-constr}

The constraints obtained from the tree--level decays are the same as
in the four--generation model, described in section
\ref{sec:the-constraints}.
The expression for $B(B\to X_s \ell^+ \ell^-)$ in the VDQ model is given by
eq. \eqref{eq:full_br_ratio_bsll}, with the
coefficients $C_2$, $\tilde{C}_9$ and $\tilde{C}_{10}$ taken as
\begin{align}
  \begin{split}
    C_2 &= 1 - \frac{U_{bs}}{\lambda_t^{bs}} ,\\
    \tilde{C}_9 &= \frac{8}{9} \ln \left (\frac{M_W}{m_B}
    \right ) \left ( 1 - \frac{U_{bs}}{\lambda_t^{bs}} \right ) + \frac{Y_0(x_t)}{\sin^2\theta_W} -
    4 Z_0(x_t) + \left (\frac{1}{\sin^2\theta_W} - 4 \right ) C_{U2Z}
    \frac{U_{bs}}{\lambda_t^{bs}},\\
    \tilde{C}_{10} &= -\frac{Y_0(x_t) + C_{U2Z} \frac{U_{bs}}{\lambda_t^{bs}}}{\sin^2\theta_W}.
  \end{split}
  \label{eq:vldt_coeff_Bsll}
\end{align}
The coefficient $C_{7 \gamma}$ remains unchanged.

The expressions for the remaining constraints are very similar to
those used in the four generation case. They are taken as in \cite{Barenboim_2001},
but with the input parameters given in tables
\ref{tab:experimental_data} and \ref{tab:theoretical_data}.
For the $\varepsilon^\prime \over \varepsilon$ constraint we use
only the ranges of the parameters given in
\cite{BURAS_SILVESTRINI_1999}. 


\subsection{Numerical results}
\label{sec:vldt_numerical-results}

\begin{figure}[htb]
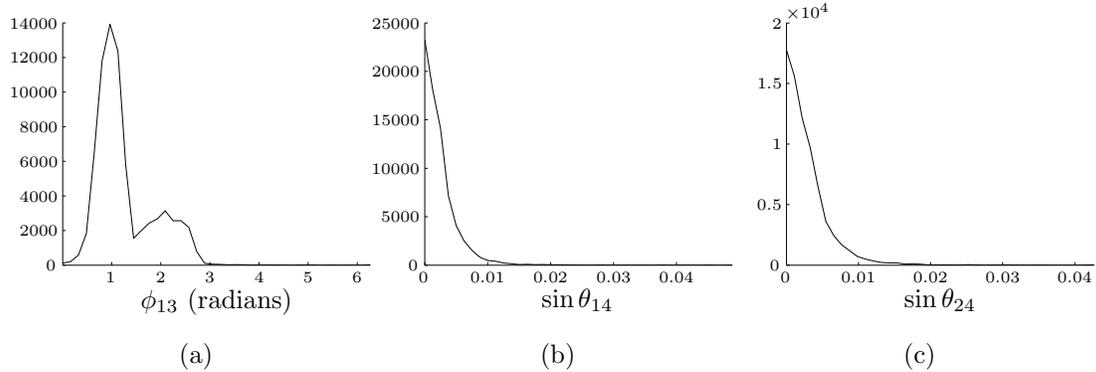

  \begin{center}
    \subfigure[]
    {
      \label{fig:phi13_vldt_hist}
      \input{vldt_phi13.tex}
    }
    \subfigure[]
    {
      \label{fig:theta14__vldt_hist}
      \input{vldt_theta14.tex}
    }
    \subfigure[]
    {
      \label{fig:theta24_vldt_hist}
      \input{vldt_theta24.tex}
    }
    \caption{Histograms of (a) $\phi_{13}$, (b) $\sin\theta_{14}$ and
    (c) $\sin\theta_{24}$ for 75,000 data points in the VDQ model}
    \label{fig:vldt_histograms}
  \end{center}
\end{figure}

The numerical scan is performed in the same way
as in the four generation case.
The results of the scan show that the CP violating phases
$\phi_{14}$ and $\phi_{24}$ can be in the entire scanned range (from 0 to $2\pi$).
Yet, as in the four generation case, the phase $\phi_{13}$
has a restricted range, $0 \lesssim \phi_{13} \lesssim
\pi$. This results agrees with \cite{Barenboim_2001}.
A histogram for this phase can be seen in figure
\ref{fig:phi13_vldt_hist}.
The mixing angles $\theta_{14}$ and $\theta_{24}$ also have restricted ranges:
\begin{align}
  \label{eq:vldt_theta_14_24_vldt_numerical_bound}
  0 \lesssim \sin\theta_{14} \lesssim 0.011\ , \nonumber\\
  0 \lesssim \sin\theta_{24} \lesssim 0.011\ .
\end{align}
The restricted regions are due to combinations of several
constraints. For example, the restricted range of
$\sin\theta_{24}$ is mainly due to the correlated effect of the
$K_L\to \mu \bar \mu$ and $\frac{\varepsilon^\prime}{\varepsilon}$
constraints. The remaining phases, $\phi_{24}$ and $\phi_{14}$ can
be in all the scanned range. The mixing angle $\theta_{34}$ can be
in the range $0 \lesssim \sin\theta_{34} \lesssim 0.12$ (see
\cite{Barenboim_2001}).

\begin{figure}[htb]
  \begin{center}
    \subfigure[]
    {
      \label{fig:vldt_NP_params_sd}
      \resizebox{4cm}{4cm}{\includegraphics{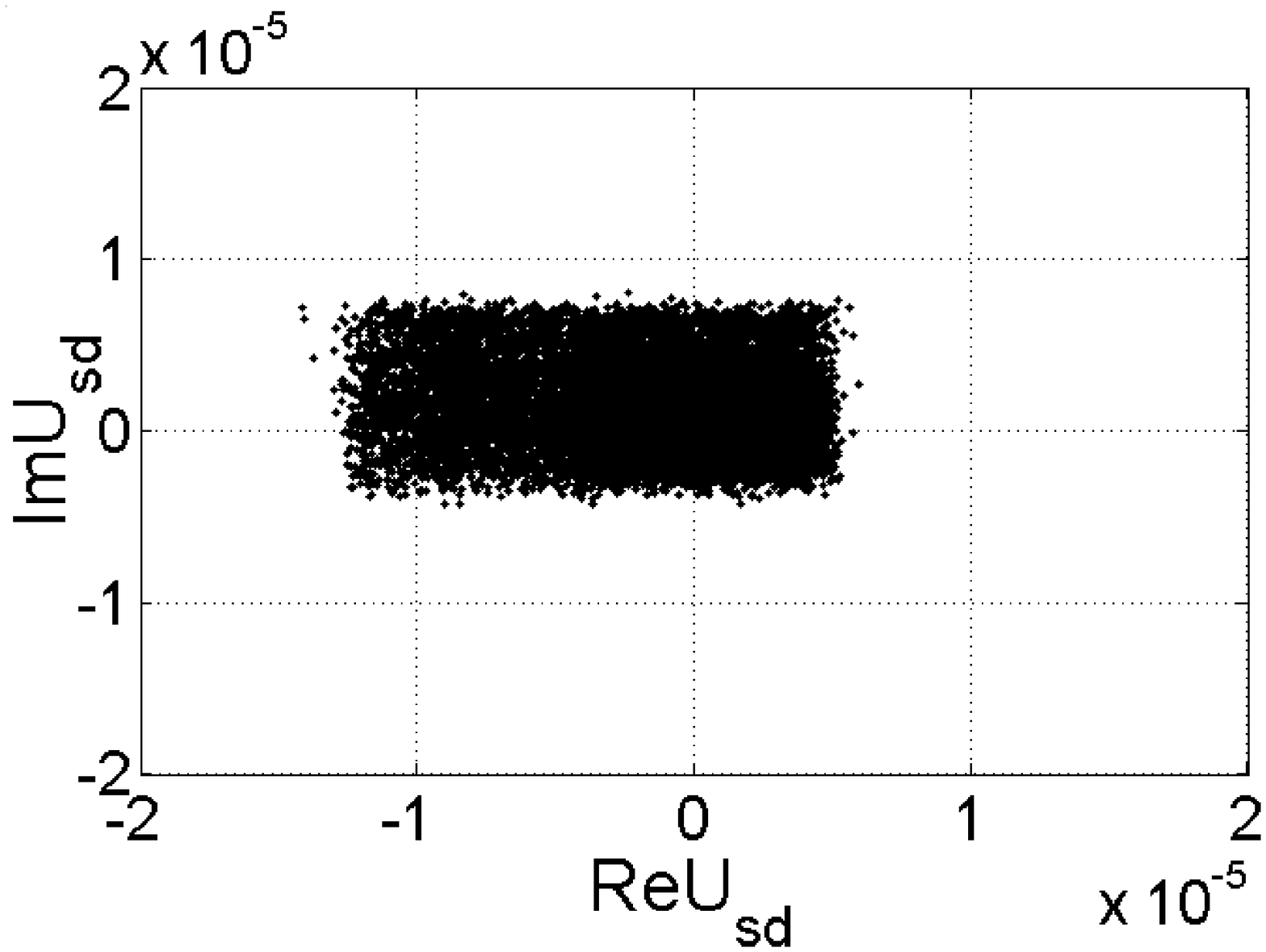}}
    }
    \subfigure[]
    {
      \label{fig:vldt_NP_params_bd}
      \resizebox{4cm}{4cm}{\includegraphics{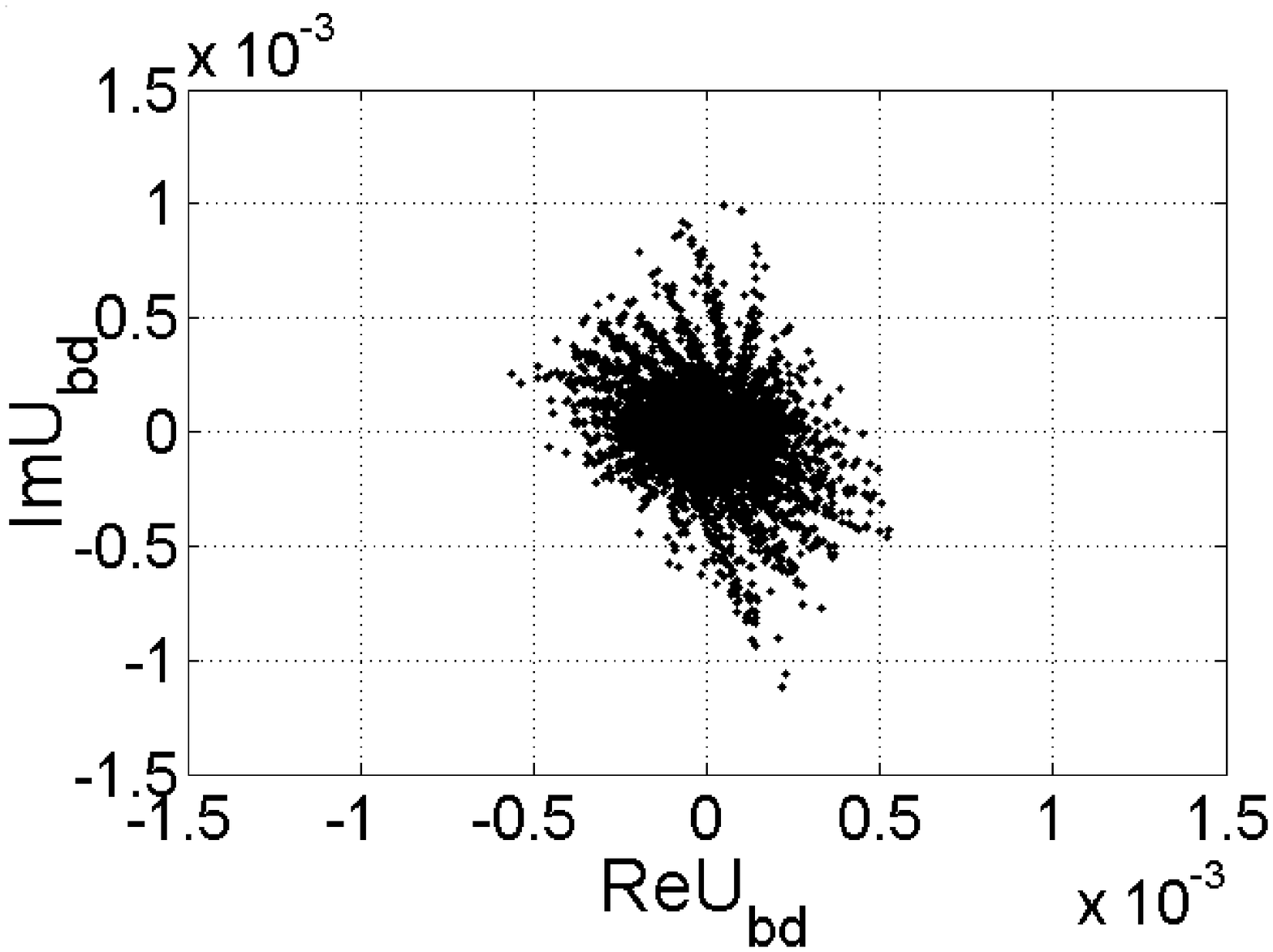}}
    }
    \subfigure[]
    {
      \label{fig:vldt_NP_params_bs}
      \resizebox{4cm}{4cm}{\includegraphics{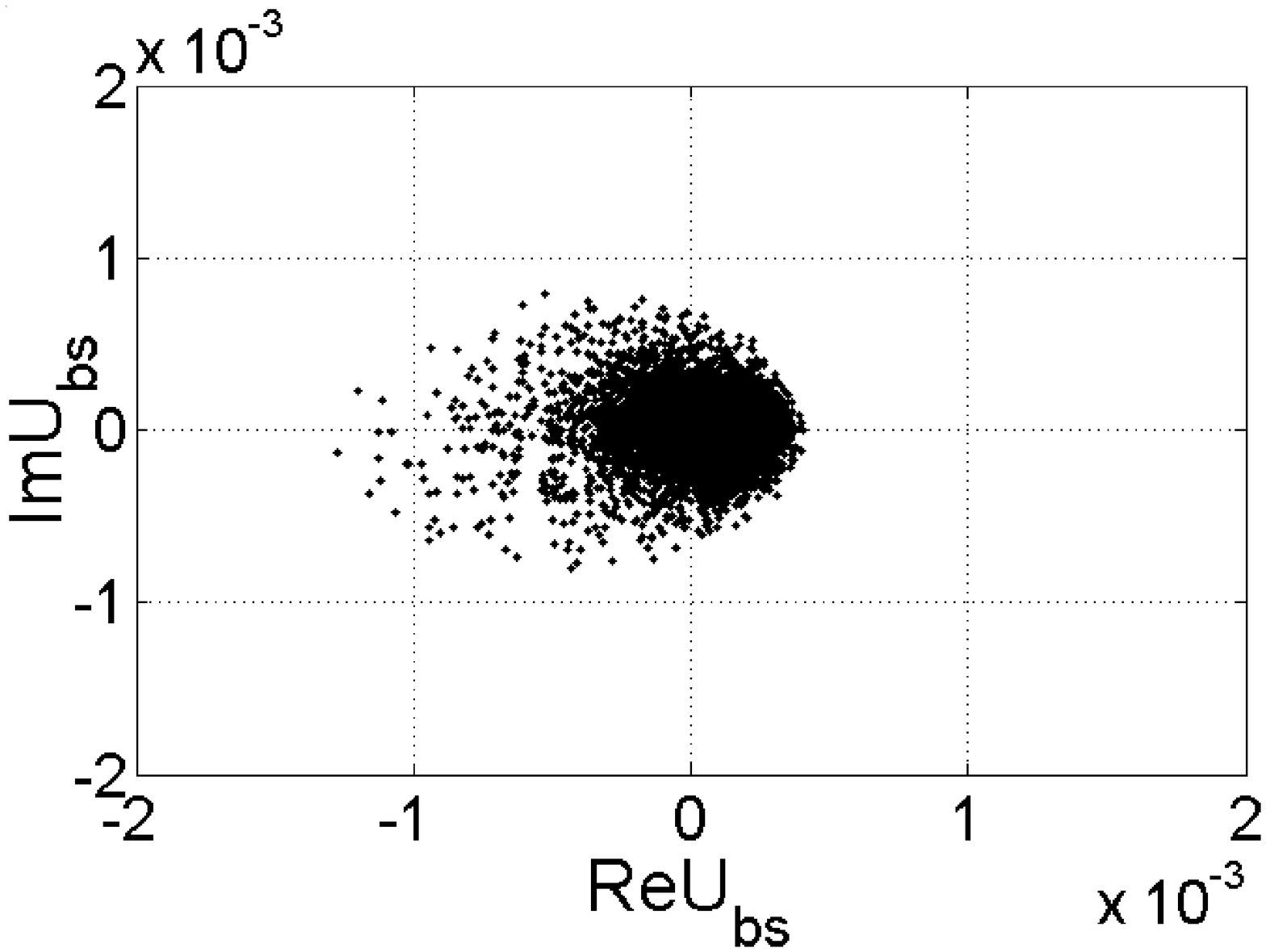}}
    }
    \caption{Scatter plot of the New Physics parameters (a) $U_{sd}$,
    (b) $U_{bd}$ and (c) $U_{bs}$ with all the constraints.}
    \label{fig:vldt_NP_params}
  \end{center}
\end{figure}

We now examine the allowed regions of the parameters
$U_{sd}$, $U_{bd}$ and $U_{bs}$, which represent the
New Physics contributions in this model. Their scatter plots are given in
Figure~\ref{fig:vldt_NP_params}. They are in very good agreement with
the results of \cite{Barenboim_2001}, except for changes which are
related to the recent $a_{\psi K}$ and $B\to X_s \ell^+ \ell^-$ measurements.
As in the case of the four generations, also in this case
each of the parameters is effectively constrained only by a small number of
constraints. We now consider each of the parameters:
\begin{enumerate}
\item
  {
    $U_{sd}$: The bound on
    $Im U_{sd}$ comes mainly from the $\varepsilon^\prime \over
    \varepsilon$ constraint.
    The real part, $Re U_{sd}$, is mainly constrained by
    $B(K_L \to \mu \bar \mu)_{SD}$.
  }
\item
  {
    $U_{bd}$: $|U_{bd}|$ is mainly constrained by the $\Delta
    m_{B_d}$ constraint.
    The $a_{\psi K}$ constraint is responsible for the area excluded in
    the upper--right and lower--left regions in figure
    \ref{fig:vldt_NP_params_bd}.
    We examined how an improvement in the experimental error for
    $a_{\psi K}$ will affect this constraint. As in the four
    generation case, the results show that the excluded area is almost
    insensitive to the size of the error, and there is little
    sensitivity also to the central value.
    This means that the coming experimental results for
    $a_{\psi K}$ are not expected to change the bounds on the flavor
    parameters of the VDQ model in a significant manner.
  }
\item
  {
    $U_{bs}$: This parameter is mainly constrained by
    the $B\to X_s \ell^+ \ell^-$ bound. It is little
    influenced even by $\Delta m_{B_s}$.
    The bounds we obtain are better than those obtained in previous works
    \cite{Barenboim_2001} due to the improved experimental bound (see
    table \ref{tab:experimental_data}) that we use.
  }
\end{enumerate}

The last step in comparing this model to the SM is to consider the
predictions given by the two models for various quantities (see
table \ref{tab:predictions1}).
We give here several notes regarding these predictions:
\begin{enumerate}
\item
  {
    The predictions for $B(K_L\to \mu \bar \mu)_{SD}$, $B(K^+\to\pi^+
    \nu \bar{\nu})$ and $B(K_L\to\pi^0 \nu \bar{\nu})$
    in the VDQ model can be substantially higher than the predictions
    of the SM. The prediction for $B(K_L\to \mu \bar \mu)_{SD}$ can
    actually span the entire range allowed by the experimental constraints.
  }
\item
  {
    The prediction for $B(B\to X_s \ell^+ \ell^-)$ in the
    VDQ model can reach the current experimental bound. This is in
    contrast to the prediction in the case of low $a_{\psi K}$
    values, as presented in \cite{Barenboim_2001}.
    In a similar way, the predictions for $B(B\to X_d \ell^+ \ell^-)$ are
    also large in this model.
  }
\item
  {
    The scan results for the CP asymmetry in semi--leptonic decays (as given in table
    \ref{tab:predictions1}) improve the bounds which were
    obtained in \cite{Eyal_1999}.
    Note that the prediction in the VDQ model
    may have a different sign than the SM prediction.
    In order for the $a_{SL}$ constraint to have a significant impact on the
    parameters of the VDQ model, the experimental results
    must improve by about an order of magnitude. Such an improvement is not expected
    in the near future.
  }
\item
  {
    The behavior of the predictions for $a_{\psi K}$ are similar to
    the four generation case. When taking all constraints at two
    sigma, the results for the SM and for the VDQ model are
    practically the same. But when performing the scan without including the $B\to \psi
    K$ constraint, the VDQ model allows the entire range for
    $a_{\psi K}$, while the SM allows only a very restricted range.
    Still, from the latest experimental results it is clear
    that the value of $a_{\psi K}$ lies at the higher
    part of the allowed range, so this information is of limited
    interest.
  }
\item
  {
    As in the four generation case, the $\Delta m_{B_s}$ prediction in the VDQ model is
    similar to the prediction of the SM. This again means that
    detection of $\Delta m_{B_s}$ outside the predictions of the SM
    will be problematic also in the VDQ model.
  }
\end{enumerate}


\section{Discussion and Conclusions}
\label{cha:discussion-conclusions}
We considered constraints from tree--level decays, electroweak
precision measurements, the decay $Z\to b \bar{b}$, rare $K$ and $B$
decays, CP violating parameters in the $K$ and in the $B$ systems and
mass differences in various neutral meson systems in order to obtain
bounds and predictions for the four generation model and for the VDQ
model.
The constraint on the four generation model from $Z\to b \bar{b}$
proves to be very powerful, as it excludes a significant portion of the
parameter space. Improvement in the $B(B\to X_s \ell^+ \ell^-)$ measurement
provides better bounds on the New Physics parameters in both models. The new
experimental data for $a_{\psi K}$ also affects our results,
causing the analysis for the VDQ model to be quite different from
the results of \cite{Barenboim_2001}. Furthermore,
according to our analysis additional improvement of this measurement
is expected to have only minor effects on both the four generation and
the VDQ models. Improvement of the bounds for these models can thus be
expected mainly by better determination of other measurements such as
$B(B\to X_s \ell^+ \ell^-)$, $B(K^+\to\pi^+ \nu \bar{\nu})$, $a_{SL}$, and $\Delta m_{B_s}$.
Of course, both of the models can never be strictly excluded by flavor
physics, since by proper adjustment of the new parameters one can
always reduce them to the SM. In the four--generation model,
this can be done by taking the mixing between the fourth generation
and the three SM generations to be very small
($\sin\theta_{34}, \sin\theta_{14}, \sin\theta_{24}\to 0$).
The result is decoupling of the fourth generation from the
other three. In the VDQ model, this can be achieved by taking the mass
of the new quark to be very high. Since the new mixing angles result
from the diagonalization of the mass matrix, they are roughly given by
$\theta_{i4} \approx \frac{m_i}{m_4}$, so by taking $m_4\gg m_i$ we
can make all the new mixing angles vanish.

We find that the two models which we consider share many
common features. They both contain new sources of CP violation that
arise from a $4\times4$ CKM-like matrix, and both predict for
various observables values that can be higher than those predicted by
the SM. Another aspect that the two models share is that they both introduce
new FCNC contributions. In both models these new
FCNC contributions are naturally small, though they can still be larger than the
SM contributions. This was shown to lead to substantial increase in
the values the two models predict for various quantities, compared to the SM
predictions. Despite these (and other) similarities between the
four--generation and the VDQ models, there are also significant
differences between them. One of the differences is related to the
fact that the mechanism that introduces the new FCNC sources is not
the same in the two models.
The new FCNC contributions in the four--generation model come from loop
diagrams with the new fermions in the loop. These diagrams are
roughly proportional to ${\left (\frac{m_{t^\prime}}{M_W} \right )}^2 \approx
O(10)$. In the VDQ model, on the other hand, the leading new FCNC contributions
come from tree--level diagrams, with a coefficient of $C_{U2Z} \approx
O(100)$ (see e.g. \cite{Barenboim_2001}). Thus the new contributions for the FCNC processes in the two
models differ by an order of magnitude.

This order--of--magnitude difference leads to various effects. One of
these regards the unitarity relation
$\lambda_u^{bd}+\lambda_c^{bd}+\lambda_t^{bd}+\lambda_{t^\prime}^{bd}=0$
in the four--generation model, or $\lambda_u^{bd}+\lambda_c^{bd}+\lambda_t^{bd}-U_{bd}=0$ in
the VDQ model.
As was already discussed in the literature (see e.g. \cite{Eyal_1999}
and references therein), in both cases one gets unitarity quadrangles instead of
the unitarity triangle that exist in the SM.
When examining the results of the numerical scan (see table \ref{tab:beta_ranges}),
it is clear that the possible shapes that the unitarity
quadrangle can take is very different in the two models.
The parameter $r$ is defined by
\begin{equation}
  \label{eq:r_definition}
  r\equiv \left\{\begin{array}{l}
      \frac{\lambda_{t^\prime}^{bd}}{\lambda_t^{bd}}
      \ \ \ \ {\rm{four\ generation\ model}}\\ \\
      \frac{U_{bd}}{\lambda_t^{bd}}
      \ \ \ \ {\rm {VDQ\ model}.} \end{array} \right.
\end{equation}
\begin{table}[htb]
\begin{center}
\footnotesize{
\begin{tabular}{|c|c|c|}
\hline
 Model & $\beta$ (degrees) & $r$ \\
\hline
\hline
 & &\\
SM & 22 - 33 & 0 \\ & &\\
four--generations ($m_{t^\prime}=500 GeV$) & 0 - 57, 281-360 &
 $\lesssim 3.5$\\ & &\\
four--generations ($m_{t^\prime}=200 GeV$) & 0 - 92, 230-360 &
 $\lesssim 70$\\ & &\\
VDQ & 2 - 38 & $\lesssim 0.16$\\ & &\\
\hline
\end{tabular} }
\caption{ \small { Allowed ranges of various parameters related to the
 unitarity quadrangle, when all constraints are taken at two
 sigma. $r$ is defined in eq. \eqref{eq:r_definition}.
}}
\label{tab:beta_ranges}
\end{center}
\end{table}
One can see that in the four--generation model,
$\lambda_{t^\prime}^{bd}$ can be significantly larger than the
corresponding SM quantity $\lambda_t^{bd}$ (up to a factor of 3.5 for
$m_{t^\prime}=500$ GeV and a factor of 70 for $m_{t^\prime}=200$ GeV),
while in the VDQ model $U_{bd}$ is only allowed to be about 15$\%$
of $\lambda_t^{bd}$. This difference can be traced back to the order--of--magnitude
difference between the new FCNC contributions in the two models. As a
result of this, also the ranges of the angle $\beta$ are significantly
different in the two models.
While in the VDQ model the unitarity quadrangle can be only slightly
modified compared to the SM triangle (due to the limited size of
r and $\beta$), in the four--generation model the shape of the
unitarity quadrangle can be completely different than that of the SM
unitarity triangle (especially in the $m_{t^\prime}=200$ GeV case).

\section*{acknowledgments}

I thank Yossi Nir for his guidance and for helpful comments on the manuscript.


\end{document}